\documentclass[twocolumn]{aastex631}
\usepackage{booktabs}
\usepackage{gensymb}
\usepackage{upgreek}
\usepackage{amssymb}

\newcommand{\kmss}{$\mathrm{km\ s}^{-1}\ $}
\newcommand{\Ms}{M$_{\odot}\ $}

\newcommand{\HIs}{\ion{H}{1} }
\newcommand{\OVIs}{\ion{O}{6} }
\newcommand{\SiIIs}{\ion{Si}{2} }

\newcommand{\SiIVs}{\ion{Si}{4} }

\newcommand{\CIVs}{\ion{C}{4} }

\newcommand{\bturbs}{$b_\mathrm{turb}\ $}
\newcommand{\sigls}{$\sigma_\mathrm{LOS}\ $}
\newcommand{\sigis}{$\sigma_\mathrm{1D}\ $}
\newcommand{\Rvirs}{$\mathrm{R}_\mathrm{vir}\ $}

\newcommand{\kms}{$\mathrm{km\ s}^{-1}$} 
\newcommand{\M}{M$_{\odot}$}

\newcommand{\OVI}{\ion{O}{6}}
\newcommand{\SiII}{\ion{Si}{2}}

\newcommand{\SiIV}{\ion{Si}{4}}

\newcommand{\CIV}{\ion{C}{4}}

\newcommand{\MgII}{\ion{Mg}{2}}
\newcommand{\bturb}{$b_\mathrm{turb}$}
\newcommand{\sigl}{$\sigma_\mathrm{LOS}$}
\newcommand{\sigi}{$\sigma_\mathrm{1D}$}

\begin{document}

\title{Constraining Circumgalactic Turbulence with QSO Absorption-line Measurements}

\shorttitle{Constraining CGM Turbulence}
\shortauthors{Koplitz et al.}

\correspondingauthor{Brad Koplitz} 
\email{brad.koplitz@asu.edu}

\author[0000-0001-5530-2872]{Brad Koplitz}
\affil{School of Earth \& Space Exploration, 
Arizona State University, 781 Terrace Mall, Tempe, AZ 85287, USA}

\author[0000-0002-6070-5868]{Edward Buie II}
\affil{Physics \& Astronomy Department, 
Vassar College, 124 Raymond Avenue, Poughkeepsie, NY 12604, USA}

\author[0000-0002-3193-1196]{Evan Scannapieco}
\affil{School of Earth \& Space Exploration, 
Arizona State University, 781 Terrace Mall, Tempe, AZ 85287, USA}

\begin{abstract}

Our knowledge of the circumgalactic medium (CGM) is mostly based on quasar absorption-line measurements. These have uncovered a multiphase medium that is likely highly turbulent, but constraints of this turbulence are limited to measurements of the non-thermal width of absorption-line components ($b_\mathrm{turb}$) and the line-of-sight velocity dispersion between components ($\sigma_\mathrm{LOS}$). Here we analyze a suite of CGM simulations to determine how well these indirect measures are related to the underlying CGM. Our simulations track the non-equilibrium evolution of all commonly observed ions, and consist of two main types: small-scale simulations of regions of homogeneous CGM turbulence and global simulations of inhomogeneous turbulence throughout a galactic halo. From each simulation, we generate mock spectra of \ion{Si}{2}, \ion{Si}{4}, \ion{C}{4}, and \ion{O}{6}, which allow us to directly compare $b_\mathrm{turb}$ and $\sigma_\mathrm{LOS}$ to the true line-of-sight turbulence ($\sigma_\mathrm{1D}$). In the small-scale simulations, $b_\mathrm{turb}$ is only weakly correlated with $\sigma_\mathrm{1D}$, likely because it measures random motions within individual warm CGM clouds, which do not sample the overall random motions. Meanwhile, $\sigma_\mathrm{LOS}$ and $\sigma_\mathrm{1D}$ are strongly correlated, with $\sigma_\mathrm{1D}\approx\sigma_\mathrm{LOS}+10\ \mathrm{km\ s}^{-1}$ in the densest regions we simulated, though, the strength of this correlation depended weakly on the gas phase being probed. Our large-scale simulations also indicate that $b_\mathrm{turb}$ and $\sigma_\mathrm{1D}$ are largely uncorrelated, and that $\sigma_\mathrm{1D}\approx\sigma_\mathrm{LOS}+10\ \mathrm{km\ s}^{-1}$ on average, although it varies along individual sightlines. Moreover, the $\sigma_\mathrm{LOS}$ distributions from our global simulations are similar to recent observations, suggesting that this quantity may provide useful constraints on circumgalactic turbulence regardless of the axis probed.

\end{abstract}


\section{Introduction}

Galaxies can only form because of the larger environment that sustains them. This circumgalactic medium (CGM) is the source of galaxies’ baryonic material, and its properties are essential to determining the history of galaxy evolution. Yet observational constraints of the structure of the CGM are extremely limited.

The primary source of this information comes from absorption-line spectroscopy, with surveys such as COS-Halos \citep{Tumlinson2013, Werk2013, Werk2016}, the Keck Baryonic Structure Survey (KBSS; \citealt{Rudie2012, Rudie2019, Erb2022}), COS-Dwarfs \citep{Bordoloi2014}, COS-GASS \citep{Borthakur2015,Borthakur2016}, and the Cosmic Ultraviolet Baryon Survey (CUBS; \citealt{Chen2020, Zahedy2021, Qu2022}), using bright background quasars as light sources. Most of this work has been at restframe UV wavelengths, though some work has been done in the X-ray \citep[e.g.,][]{Comparat2022, Nicastro2023}. Observations such as these have shown that the CGM is highly ionized, extended, and multiphase, with warm ions such as \CIVs and \SiIVs being found at similar velocities as colder ions, such as \MgII, \ion{N}{2}, and \SiIIs \citep{Tripp2011, Meiring2013, Burchett2015, Burchett2019}. Furthermore, this cold gas is often associated with gas in higher ionization states including \OVIs and \ion{Ne}{8}, which usually occur at physical conditions that indicate rapid cooling. 

Together, these observations directly imply the presence of radically different phases in close spatial proximity and the rapid cycling of baryons amongst these phases \citep[e.g.][]{Lochhaas2020}. Thus, galaxy formation must be understood as embedded within a baryon cycle in which: (i) material continually cools and flows onto galaxies \citep[e.g.,][]{Keres2005, Dekel2006, Keres2009, Stewart2011, Rubin2012, Voit2015}, (ii) supernova-driven outflows put significant mass and energy back into the to the CGM \citep[e.g.,][]{Heckman1990, Martin2005, Veilleux2005, Steidel2010, Martin2012, Scannapieco2015, Thompson2016, Scannapieco2017, Chisholm2017, Fielding2007}, and (iii) accreting black holes act as even more efficient energy sources, ejecting material into the CGM in the form of wide-angle winds and relativistic jets \citep{Scannapieco2004, DiMatteo2005, Croton2006, Thacker2006, McNamara2007, Moe2009, Hamann2011, Fabian2012, Arav2013}.

All of these processes drive significant turbulence, which is likely to exist at levels that alter the properties of the CGM. In fact, the CGM is particularly sensitive to turbulent motions, as rapid cooling lowers the temperature of much of the medium to significantly below the virial temperature. This means that random motions on the order of $\approx$0.25 the virial velocity, as observed in galaxy groups and clusters \citep[e.g.,][]{Werner2009, Hitomi2016, Ogorzalek2017, Zhuravleva2018, Eckert2019}, can give rise to supersonic motions in the CGM. The resulting shocks can produce density and temperature structures that are dependent on the average turbulent Mach number \citep{Buie2020a}, strongly affecting the baryon cycle and its impact on galaxy formation.

Accurately capturing these processes in cosmological simulations presents two major challenges. First, simulating the CGM at sufficient resolution to follow turbulent motions in a cosmological setting can be very computationally expensive \citep[e.g.,][]{Hummels2019,Peeples2019,vandeVoort2019,Lochhaas2022}. Second, the large uncertainties in the turbulent properties of the CGM make it difficult to sufficiently span the parameter space allowed by current observations.

This uncertainty arises because, unlike in galaxy clusters, the scale at which turbulence is probed by observations may differ substantially from the scale at which it is driven. CGM measurements of turbulence consist of two main types: (i) the turbulent broadening associated with the line profile of individual absorbers, and (ii) the distribution of the velocity centroids of absorption-line systems relative to one another. While both of these types of measurements provide constraints on random velocities, these are also both likely to underestimate the total level of turbulence. 
This is due to the turbulent cascade causing the velocities to be distributed as a function of scale, as $v (L) \propto ({L}/L_{\rm drive})^\alpha$, where $1/3 \leq \alpha \leq 1/2$ depends on the Mach number of the medium \citep{Kritsuk2007, Pan2010}. As a result, understanding how well observational tracers recover the turbulent velocities requires a careful comparison between observations and simulations. 

Motivated by these issues, we undertake a comprehensive investigation to quantify how observational measurements of turbulence are related to the underlying CGM kinematics. Our study builds on the work of \citet{Buie2018, Buie2020b, Buie2022}, who carried out a large suite of idealized simulations to model the impact of turbulence on the properties of the CGM. 
These were of two types: (i) small-scale simulations of regions of homogenous CGM turbulence, and (ii) global simulations of inhomogenous turbulence that was driven in a galactic halo with gravitational acceleration and co-rotation. Here we make use of both types of simulations to investigate the motions inferred from the two most common estimators of turbulence and compare them to the true values of the random motion in these simulations.

The structure of this paper is as follows: \S \ref{sect:maihem} introduces the simulations we analyze, \S \ref{sect:mockobs} shows how we generate spectra from sightlines through the simulations, and \S \ref{sect:turb} details how we estimate the amount of turbulence present in individual components and along sightlines. \S \ref{sect:results} presents the results of our analysis, quantifying the relationship between true and inferred turbulent motions. A summary and conclusions are given in \S \ref{sect:summary}.

\section{MAIHEM Simulations} \label{sect:maihem}

The simulations we analyze make use of the open-source Models of Agitated and Illuminated Hindering and Emitting Media (MAIHEM\footnote{\url{http://maihem.asu.edu/}}) code to model turbulence in the CGM. MAIHEM is a three-dimensional (3D) cooling and chemistry package built using FLASH (Version 4.3; \citealt{Fryxell2000}). 
The version we used tracks the reaction network of 65 ions: hydrogen (\HIs and \ion{H}{2}), helium (\ion{He}{1} $–$ \ion{He}{3}), carbon (\ion{C}{1} $–$ \ion{C}{6}), nitrogen (\ion{N}{1} $–$ \ion{N}{7}), oxygen (\ion{O}{1} $–$ \ion{O}{8}), neon (\ion{Ne}{1} $–$ \ion{Ne}{10}), sodium (\ion{Na}{1} $–$ \ion{Na}{3}), magnesium (\ion{Mg}{1} $–$ \ion{Mg}{4}), silicon (\ion{Si}{1} $–$ \ion{Si}{6}), sulfur (\ion{S}{1} $–$ \ion{S}{5}), calcium (\ion{Ca}{1} $–$ \ion{Ca}{5}), iron (\ion{Fe}{1} $–$ \ion{Fe}{5}), and electrons. 
For each of these species, the code tracks collisional ionizations by electrons, photoionizations by a UV background, dielectric and radiative recombinations, and charge transfer reactions. 
MAIHEM was developed in \citet{Gray2015} and later expanded upon in \citet{Gray2016} and \citet{Gray2017}. 
The simulations analyzed in this work were run with a 0.3 $Z_\odot$ metallicity and exposed to a \citet{HM2012} extragalactic UV background at $z = 0$.

\subsection{Small-Scale Simulations}

To understand turbulence in the CGM on small scales, we analyze the homogenous box simulations by \citet{Buie2018}.
Turbulence is driven in these simulations through the use of an artificial forcing term $\bf{F}$ incorporated into the momentum equation as
\begin{equation}
\frac{\partial{\rho \bf{v}}}{\partial{t}} + \nabla (\rho \mathbf{v}\mathbf{v}) + \nabla P = \rho \mathbf{F},
\end{equation}
where $\rho$ is the density, $P$ is the pressure, and $\bf{v}$ is the velocity. 
The forcing term was modeled as a stochastic Ornstein-Uhlenbeck process \citep{Uhlenbeck1930, Schmidt2009, Federrath2010, Pan2010} with a user-defined forcing correlation time $t_f$, approximately equal to the eddy turnover time. 
For all the simulations presented here, turbulence was driven solely through solenoidal modes (i.e., $\nabla \cdot {\bf F} = 0$) in the range of wavenumbers 1 $\le L_{\rm box} |{\bf k}|/2 \pi \le$ 3, such that the average forcing wavenumber was $k_f^{-1} \simeq 2 L_{\rm box}/2 \pi,$ with $L_{\rm box}$ the size of our turbulent box, which was fixed at 100 parsecs on a side. 
This turbulence was continuously driven throughout the simulation runtime. 

The parameter space spanned by the simulations is greatly simplified by the dependencies of turbulent decay and cooling on density and length scale. In particular, the simulations are invariant under transformations in which $x \rightarrow \lambda x$, $t \rightarrow \lambda t$, and $\rho \rightarrow \rho/\lambda$. This means that the species fractions and thermal state of the gas will only depend on the ionization parameter ($U$), \sigi, and the product of the mean density and the turbulent driving scale, $nL_\mathrm{drive}$. Here $U \equiv \frac{\Phi}{n_\mathrm{H} c}$, where $\Phi$ is the total flux of ionizing photons, $n_\mathrm{H}$ is the total hydrogen density, and $c$ is the speed of light. The simulations we analyze have log $U$ between $-3$ and $-1$ which matches the observations of the COS-Halos survey \citep{Werk2014}.

We considered turbulence driven to one-dimensional velocity dispersions of \sigis = 6, 17, 26, 35, 46, and 58 \kmss and $n$ = 0.1, 1, and 10 cm$^{-3}$. With the range of driving modes described above, each simulation has an average driving scale of $L_{\rm drive} \approx 40$ pc leading to $nL_{\rm drive} \approx 10^{19}$, $10^{20}$, and $10^{21}\ \mathrm{cm}^{-2}$. 
Each simulation was started with a uniform density and temperature $T = 10^5$ K, which is roughly the density and temperature of various phases within the CGM \citep{Tumlinson2017}.
These simulations have a resolution of 128$^3$, corresponding to 0.8 pc, and they were evolved until reaching a steady state solution. As a result, the duration of each run varies.

\begin{figure*}
    \centering
    \includegraphics[width=\linewidth]{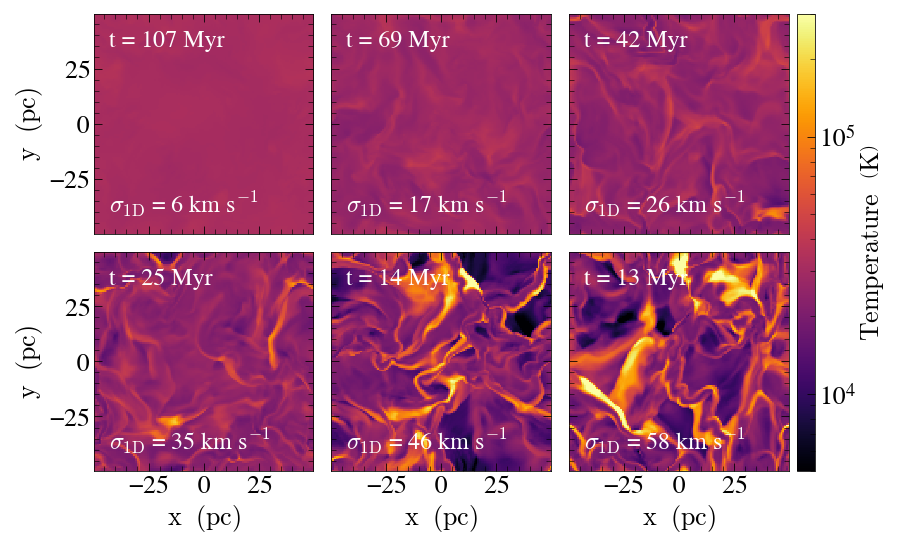}
    \caption{Temperature slices in the $xy-$plane through the $nL_\mathrm{drive} = 10^{20}\ \mathrm{cm}^{-2}$ density small-scale simulations. The injected \sigis is shown in the lower left of each panel. The upper left shows the amount of time each simulation was ran for.}
    \label{fig:small_temp_slices}
\end{figure*}

In Figure \ref{fig:small_temp_slices}, we show temperature slices through the $nL_\mathrm{drive} = 10^{20}$ cm$^{-2}$ boxes as an example of how the small-scale simulations look. 
At lower turbulent velocities (\sigis $\leq$ 26 \kms), the gas is subsonic, yielding weak shocks and small temperature gradients. 
At higher velocity dispersions, we begin to see larger temperature gradients caused by stronger shocks. These sweep across cells carrying newly-ionized material in their wakes and lead to a large spread in the range of species formed in the simulations. 
Note also how the structures change from more cloud-like at low Mach numbers, to more planar at high Mach numbers, as $\alpha$, the slope of the turbulent cascade becomes steeper and the fractal dimension of the velocity field changes \citep[e.g.,][]{Pan2011}.

\begin{figure*}
    \centering
    \includegraphics[width=\linewidth]{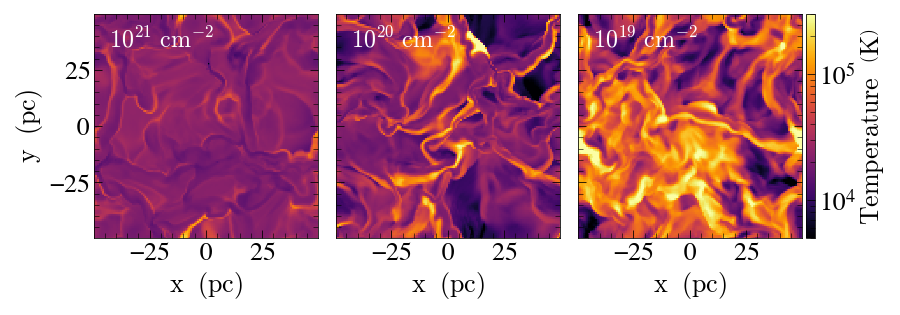}
    \caption{Temperature slices in the $xy-$plane through each simulation with \sigis $= 46$ \kms. The upper left of each panel shows the $nL_\mathrm{drive}$ of the simulation being shown.}
    \label{fig:small_den_slices}
\end{figure*}

In Figure \ref{fig:small_den_slices}, we show temperature slices through each box with \sigis $= 46$ \kmss to indicate how changing $nL_\mathrm{drive}$ impacts the simulated gas. By decreasing $nL_\mathrm{drive}$, we reduce the eddy turnover time, which increases the energy input rate and the temperature. Thus, decreasing decreasing $nL_\mathrm{drive},$ produces a higher level of ionization in the gas.

A detailed discussion of all the small-scale simulations used in this study can be found in \citet{Buie2018}. Additional simulations were also run by \citet{Buie2018} at lower densities, but these either: (i) failed to achieve a steady state, leading to thermal runaway, or (ii) failed to produce $N$ above our cutoff of $10^{11}$ cm$^{-2}$, as discussed below in \S \ref{sect:voigt}. 
As a result, we exclude them from our analysis.

\begin{figure*}
    \centering
    \includegraphics[width=\linewidth]{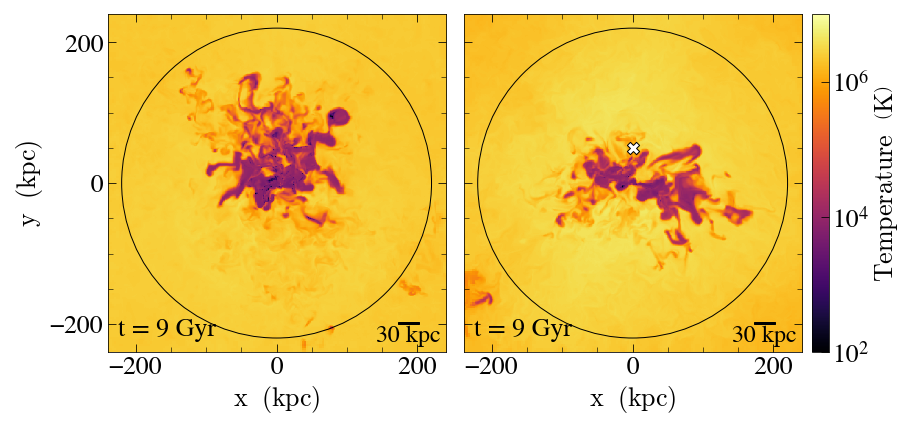}
    \caption{Temperature slices in the $xy-$plane through the Hydro (left) and MHD (right) simulations after 9 Gyr of evolution. The virial radius of the simulation is shown by a black circle. The white X shows the position of the example sightline shown in Figure \ref{fig:ex_sightline}. Each simulation ran for 9 Gyr, which we show in the lower left of each panel.}
    \label{fig:large_temp_slice}
\end{figure*}

\subsection{Large-Scale Simulations}

We also consider two large-scale simulations from \citet{Buie2022}. Each was run with the same initial conditions except that one followed the equations of magnetohydrodynamics, which we refer to as the MHD run, and one did not include magnetic fields, which we refer to as the Hydro run. These simulations were carried out in (800 kpc)$^3$ sized boxes with turbulence injected as a function of radius from the simulation center such that
\begin{equation}\label{eq_turb}
    a_{x,y,z} = a^0_{x,y,z} \left(\frac{r + 0.3\ \mathrm{R}_\mathrm{vir}}{0.5\ \mathrm{R}_\mathrm{vir}} \right)^{-1},
\end{equation}
where $a^0_{x,y,z}$ is the central acceleration and \Rvirs is the virial radius of the simulated galaxy.

Additionally, the initial gas density was chosen to follow an \citet[][NFW]{Navarro1996} profile with a dark matter halo described by
\begin{equation}
    \rho(r) = \frac{\rho_0}{\frac{r}{\mathrm{R}_s} \left(1 + \frac{r}{\mathrm{R}_s}\right)^2 },
\end{equation}
where
 $   \rho_0 \equiv {\mathrm{M}_\mathrm{halo}}\left\{4\pi \mathrm{R}_s^3\left[\mathrm{ln}(1+c)-{c}({1+c})^{-1}\right]\right\}^{-1}$
is the central dark matter density, R$_s = \mathrm{R}_\mathrm{vir}/c$ is the scale radius, and $c$ and M$_\mathrm{halo}$ are the concentration parameter and halo mass. 
The simulations we investigate here were run for 9 Gyr with \Rvirs = 220 kpc, $c = 10$, and M$_\mathrm{halo} = 10^{12}$ M$_\odot$. The virial temperature in this case is $\approx$1.2 $\times 10^6$ K, corresponding to a sound speed of $\approx$ 170 \kms. 

The $L_\mathrm{drive}$ of the simulations were chosen to be 30 kpc, which is approximately the size of the Milky Way disk. This led to $nL_\mathrm{drive}$ values which are comparable to the small-scale simulations.
Additionally, each cell in the simulations was exposed to the same extragalactic UV background \citep{HM2012}. This, combined with the initial density profiles described above, led to the initial ionization fraction varying from $10^{-8}$ in the simulated galaxy, to $10^{-5}$ just outside the galaxy, and 0.9 in the outer regions of the CGM. A refined mesh grid was used for these large-scale simulations. Our sightlines probe these boxes within \Rvirs of the center, where the resolution is highest with 512$^3$, corresponding to 1.6 kpc.

The magnitude of the turbulent driving was set such that the initial average turbulent motion is $\approx$45 \kmss within the virial radius, but additional convective motions are generated as the simulation evolves, increasing the overall turbulence at late times.
In both simulations, we initialized the medium with a Keplerian circular velocity modified by the observational findings from the Mg II studies of \citet{Ho2017}, using their Equation (A2) to set how velocities should fall off along the minor axis, which is chosen to be the $y$-axis in our runs. 
This gives a rotational velocity of 
\begin{equation} 
    v_{\rm rot} = f_{\rm rot} \sqrt{\frac{4 \pi G r^2 \rho_{\rm NFW}(r)}{3}} H_s(y),
    \label{equ:rotation}
\end{equation}
where $G$ is the gravitational constant, $H_s~\equiv~\exp{(-|y|/{\rm 50~kpc})}$, and $f_{\rm rot}$ is an overall scaling parameter set to 60\%. In the MHD case, the simulation included a uniform seed magnetic field in the $z$-direction of 0.1$~\mu$G run.
We refer the reader to \citet{Buie2020b} and \citet{Buie2022} for additional details on the simulations.

We analyzed each simulation after 9 Gyr of evolution, and measured the turbulent profile in detail as discussed in \S \ref{sect:large}.
In Figure \ref{fig:large_temp_slice}, we show temperature slices in the $xy-$plane through the Hydro (left) and MHD (right) simulations. At this time, most of the volume is filled with a diffuse ambient medium and dense structures, such as clouds and filaments. The Hydro run tends to have a slightly wider range of temperatures, such that a significant population of cold clumps can be seen at temperatures down to a few hundred degrees. In the MHD simulation, however, the temperature range is more limited, likely due to the presence of magnetic pressure, which puffs up the clumps reducing their overall cooling. 
At the same time, magnetic fields aid in the transport of angular momentum, such that the overall population of dense $10^2-10^4$ K gas is closer to the center in the MHD run.

\section{Generating Mock Observations} \label{sect:mockobs}

Our simulated absorption-line observations are generated by calculating the optical depth as a function of frequency ($\nu$) in each grid cell along the sightlines as
\begin{equation}\label{eq_tau}
    \uptau_\nu = \int n_X x^{n+} \sigma_\nu \mathrm{d}\nu,
\end{equation}
where $n_X$ is the number density of element $X$ and $x^{n+}$ is the fraction of element $X$ atoms in the $n$th ionization state. 
We calculate $\uptau_\nu$ in each cell along the sightline, summing all cells together to get the total optical depth. Here the absorption cross section ($\sigma_\nu$) can be expressed as
\begin{equation}
    \sigma_\nu = \frac{\pi e^2}{m_e c} f_{lu} \phi_\nu,
\end{equation}
where $e$ and $m_e$ are the charge and mass of an electron, respectively, $f_{lu}$ is the oscillator strength of the transition \citep{Morton2003}, and the line profile ($\phi_\nu$) is a Gaussian of the form:
\begin{equation}
    \phi_\nu = \frac{1}{\Delta\nu_D \sqrt{\pi}} \mathrm{exp}\left[- \left(\frac{\nu - \nu_{lu}}{\Delta\nu_D}\right)^2 \right],
\end{equation}
where
\begin{equation}
    \Delta\nu_D \equiv \frac{\nu_{lu}}{c}b_X = \frac{\nu_{lu}}{c} \sqrt{\frac{2k_BT}{m_X}},
\end{equation} 
$k_B$ is the Boltzmann constant, and $m_X$ and $b_X$ are the mass and Doppler width of element $X$, respectively.

\begin{figure}
    \centering
    \includegraphics[width=\linewidth]{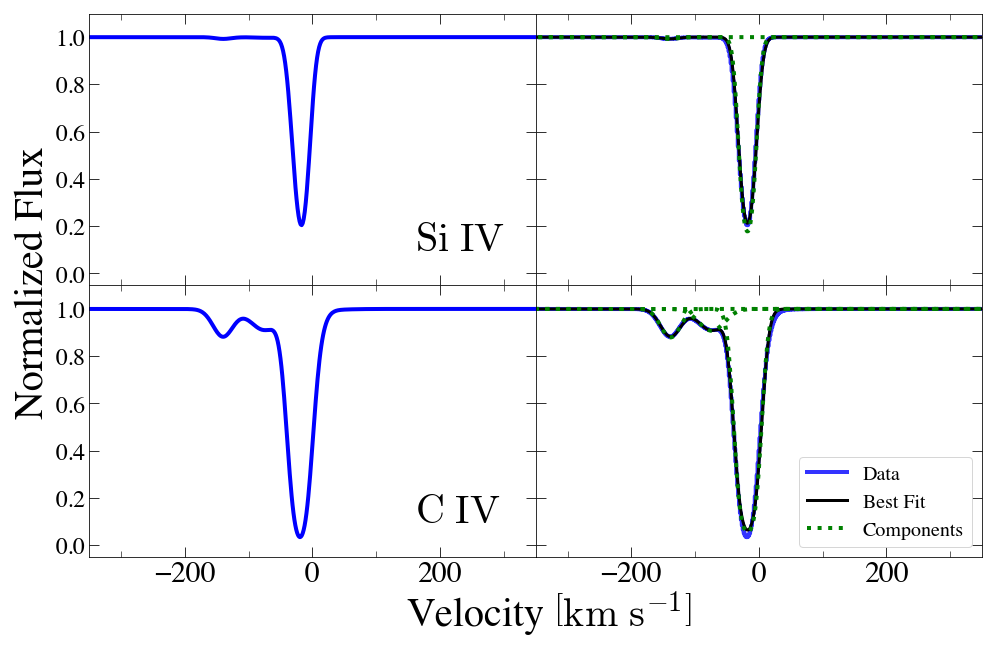}
    \caption{Example spectrum of \SiIVs (top) and \CIVs (bottom) through the large-scale MHD simulation with $\rho = 50$ kpc is shown in the left panel. The position of this example sightline is shown in Figure \ref{fig:large_temp_slice} as a white X. The resulting Voigt profile fits from \texttt{RBVFit} are shown in the right panels.}
    \label{fig:ex_sightline}
\end{figure}

To simulate observations taken with Cosmic Origins Spectrograph (COS) aboard the Hubble Space Telescope (HST), we convolve the resulting spectrum with an 8 \kmss wide Gaussian profile \citep{Osterman2011}. 
We show an example sightline through the large-scale MHD box in the left panel of Figure \ref{fig:ex_sightline}, which probes the halo at an impact parameter of $\rho = 50$ kpc at the position indicated in Figure \ref{fig:large_temp_slice} with a white X.

\section{Inferring Turbulent Velocities} \label{sect:turb}

\subsection{Fitting Voigt Profiles} \label{sect:voigt}

Once the absorption lines have been generated, we fit them to Voigt profiles using the Bayesian fitting code \texttt{RBVFit}\footnote{\url{https://github.com/rongmon/rbvfit}}, which takes in a normalized spectrum and returns the best-fit column density ($N$), Doppler width parameter ($b$), and velocity centroid ($v$) of each absorption feature along with the associated uncertainties. 
An example of these fits is shown in the right panels of Figure \ref{fig:ex_sightline}.

We limit our analysis to components with $N \geq 10^{11}$ cm$^{-2}$.
This limit was chosen to be below the observational limit of COS ($N \approx 10^{12}$ cm$^{-2}$; \citealt{Tumlinson2017}) and above {$N \approx 10^{10}$ cm$^{-2}$, where the uncertainties became so large that the returned values were unreliable.

While all of our simulated ions have multiple UV transitions, we only consider the stronger line of each ion. 
Each line will have the same shape but differ by the depth of the absorption feature according to the relative oscillator strengths. 
Therefore, little information can be gained from generating and fitting all lines in a transition.

\begin{figure*}
    \centering
    \includegraphics[width=\linewidth]{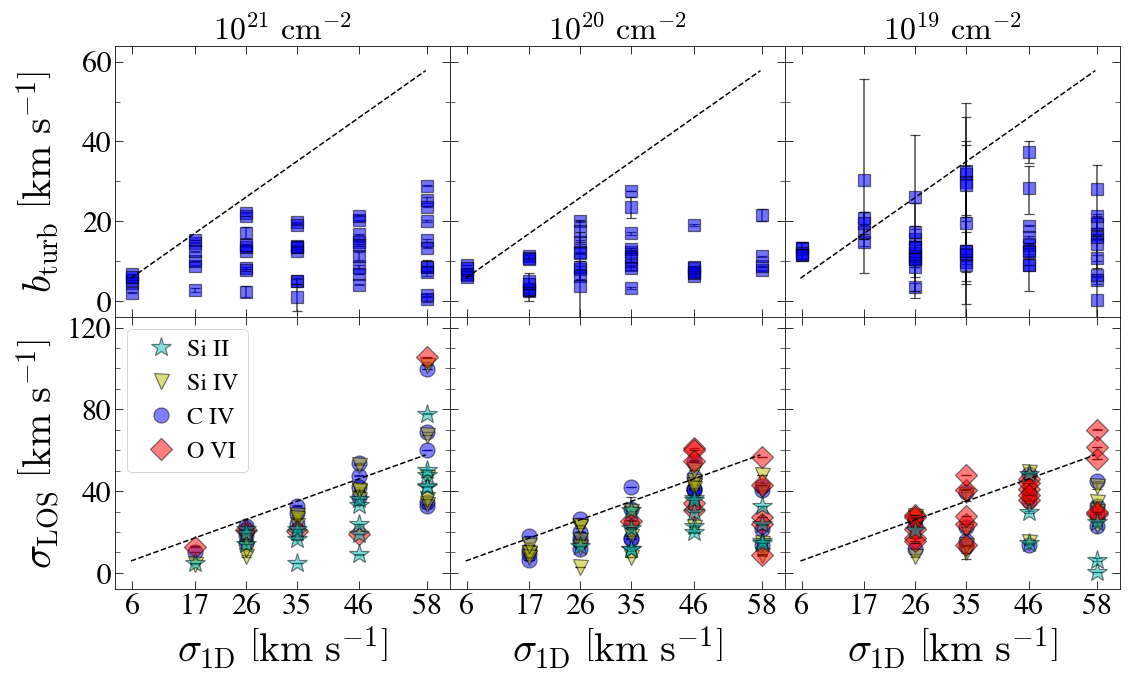}
    \caption{Turbulent measures in the small-scale simulations as a function of injected \sigi, where the top panels show \bturbs and the bottom panels are \sigl. Each column is labeled with the $nL_\mathrm{drive}$ of the box the results are through. The coloring and symbol in the top panel indicates the ion being shown with \CIVs being blue circles, yellow triangles for \SiIV, \OVIs are red diamonds, and \SiIIs shown as cyan stars. The dashed black line in each panel shows the 1$-$to$-$1 line in each panel.}
     \vspace{0.05in}
    \label{fig:res1D}
\end{figure*}

\subsection{Turbulent Velocity Measures} \label{sect:def_turb}

Our Voigt profile fits allow us to measure turbulence in multiple ways. 
With the Doppler $b$ parameters of \CIVs and \SiIV, we are able to determine the turbulence in individual components while the $v$ of each ion can be used to find the turbulence along the entire sightline. 
Since each measure uses different parameters, they are independent tracers of the turbulence.

Following \citet{Rauch1996}, we use \CIVs ($\lambda\lambda$1548, 1550 \AA) and \SiIVs ($\lambda\lambda$1393, 1402 \AA) to separate the observed line profiles into thermal and turbulent or ``non-thermal'' components. 
To do this we assume that $b$ values can be expressed as a thermal component ($b_\mathrm{therm}$) and a non-thermal or turbulent component (\bturb):
\begin{equation}
    b_X^2 = b^2_{\mathrm{therm,}X} + b^2_{\mathrm{turb}} = \frac{2k_B T}{m_X} + b^2_{\mathrm{turb}},
\end{equation}
where $X$ is either \CIVs or \SiIVs for our analysis. 
These equations can be rearranged to give:
\begin{equation}\label{eq:bturb}
    b^2_{\mathrm{turb}} = \sqrt{b^2_\mathrm{CIV} - b^2_\mathrm{therm,CIV}},
\end{equation}
and
\begin{equation}\label{thermal}
    b^2_{\mathrm{therm,CIV}} = \frac{b^2_{\mathrm{CIV}} - b^2_{\mathrm{SiIV}}}{1 - (m_\mathrm{C} / m_\mathrm{Si})},
\end{equation}
which allows us to deconvolve $b_\mathrm{therm}$ from \bturbs of the Doppler parameter and determine the turbulence in each component. 
This measure has been used to constrain the turbulence in individual absorbers along quasar sightlines (e.g., \citealt{Rauch1996, Rudie2019}).

The turbulence along the entire sightline can be probed by measuring the line-of-sight velocity dispersion, \sigls is defined as
\begin{equation}\label{eq:sig_los}
    \sigma_\mathrm{LOS}^2 = \langle \left[v - \langle v\rangle \right]^2 \rangle,
\end{equation}
where the angled brackets indicate taking the mean. 
\sigls has been used to observationally constrain the amount of turbulence along quasar sightlines (e.g., \citealt{Zhu2014, Borthakur2016, Huang2016}). 
In addition to the warm transitions of \CIVs and \SiIV, we generate spectra of cool (\SiII; $\lambda\lambda$1260, 1193, 1190, 1304 \AA) and hot (\OVI; $\lambda\lambda$1032, 1037 \AA) transitions, also using them to measure \sigl.
This allows us to quantify how turbulence impacts the different phases of the CGM.

Currently, observers use \bturbs and \sigls to constrain the amount of turbulence in the universe; however, the robustness of the methods have not been well tested. To that end, we investigate these measures in MAIHEM simulations in which the ``true'' turbulent velocities are known.

\begin{deluxetable*}{CCCCCCC}[t]
    \label{tab:sig1D_results}
    \centering
    \tablecaption{Mean turbulence results through the small-scale simulations.}
    \tablehead{
     \colhead{$\sigma_\mathrm{1D}$} & \colhead{\bturb} & \colhead{$b_\mathrm{therm}$} & \colhead{\sigls (\SiII)} & \colhead{\sigls (\SiIV)} & \colhead{\sigls (\CIV)} & \colhead{\sigls (\OVI)}\\
    \colhead{$\left[\mathrm{km\ s}^{-1}\right]$} & \colhead{$\left[\mathrm{km\ s}^{-1}\right]$} & \colhead{$\left[\mathrm{km\ s}^{-1}\right]$} & \colhead{$\left[\mathrm{km\ s}^{-1}\right]$} & \colhead{$\left[\mathrm{km\ s}^{-1}\right]$} & \colhead{$\left[\mathrm{km\ s}^{-1}\right]$} & \colhead{$\left[\mathrm{km\ s}^{-1}\right]$}}
    \startdata
    \multicolumn{7}{c}{$nL_{\rm drive} = 10^{21}$ cm$^{-2}$ Simulations} \\
    \hline
    6 & 5.1 \pm 1.8 & 13.7 \pm 0.4 & --- & --- & --- & --- \\
17 & 10.7 \pm 4.6 & 12.6 \pm 0.6 & 4.6 & 4.4 & 10.1 & 12.7 \\
26 & 13.3 \pm 6.4 & 12.5 \pm 5.5 & 18.2 \pm 3.8 & 13.5 \pm 4.3 & 19.2 \pm 3.5 & 20.8 \\
35 & 11.5 \pm 6.5 & 13.3 \pm 3.7 & 15.6 \pm 7.6 & 26.1 \pm 2.3 & 28.7 \pm 4.6 & 20.3 \\
46 & 13.2 \pm 5.9 & 14.5 \pm 4.1 & 24.2 \pm 10.9 & 42.4 \pm 7.4 & 45.0 \pm 7.2 & 19.0 \\
58 & 14.0 \pm 9.6 & 10.2 \pm 3.9 & 52.2 \pm 14.8 & 57.2 \pm 28.2 & 59.2 \pm 27.7 & 105.4 \\
\hline
\multicolumn{7}{c}{$nL_{\rm drive} = 10^{20}$ cm$^{-2}$ Simulations} \\
\hline
6 & 7.4 \pm 1.3 & 10.6 \pm 0.3 & --- & --- & --- & --- \\
17 & 6.4 \pm 3.9 & 11.4 \pm 3.5 & --- & 10.2 \pm 2.6 & 11.1 \pm 4.9 & --- \\
26 & 11.5 \pm 5.4 & 10.5 \pm 1.8 & 13.1 & 15.4 \pm 8.5 & 18.1 \pm 5.5 & --- \\
35 & 13.9 \pm 7.7 & 11.3 \pm 3.8 & 18.9 \pm 9.1 & 21.4 \pm 9.1 & 26.2 \pm 10.6 & 25.4 \\
46 & 9.4 \pm 4.8 & 11.4 \pm 1.5 & 28.9 \pm 7.5 & 40.3 \pm 10.5 & 44.7 \pm 6.0 & 48.2 \pm 14.6 \\
58 & 12.4 \pm 6.4 & 11.0 \pm 0.4 & 21.6 \pm 8.6 & 27.9 \pm 19.1 & 29.4 \pm 14.3 & 32.0 \pm 18.4 \\
\hline
\multicolumn{7}{c}{$nL_{\rm drive} = 10^{19}$ cm$^{-2}$ Simulations} \\
\hline
6 & 12.3 \pm 0.9 & 6.1 \pm 1.9 & --- & --- & --- & --- \\
17 & 20.6 \pm 5.9 & 7.6 \pm 3.1 & --- & --- & --- & --- \\
26 & 13.0 \pm 5.7 & 10.8 \pm 3.3 & 20.8 & 21.6 \pm 9.1 & 22.0 \pm 7.0 & 22.0 \pm 5.7 \\
35 & 19.9 \pm 9.9 & 12.9 \pm 3.9 & --- & 20.8 \pm 12.3 & 22.6 \pm 12.0 & 30.7 \pm 13.7 \\
46 & 17.6 \pm 9.5 & 10.9 \pm 9.7 & 30.9 \pm 17.1 & 34.7 \pm 14.5 & 34.0 \pm 14.7 & 40.9 \pm 4.2 \\
58 & 13.9 \pm 8.3 & 13.2 \pm 7.2 & 10.6 \pm 12.9 & 32.7 \pm 8.0 & 32.1 \pm 9.3 & 49.2 \pm 18.9 \\
    \enddata
    \tablenotetext{}{The uncertainties show the standard deviation on the distribution. Simulations for which \sigls could only be measured along one sightline do not have an associated standard deviation. Blank entries show simulations where we were unable to measure \sigls along any sightline.}
\end{deluxetable*}
 
\section{Results \& Discussion} \label{sect:results}

\subsection{Small-Scale} \label{sect:small}

The mean and spread of our turbulent measures through each small-scale simulation are given in Table \ref{tab:sig1D_results}. 
Simulations without \sigls values indicate that only one component was seen in each sightline and so \sigls could not be measured. Similarly, if we could only measure \sigls for one sightline, we do not report a spread. In each of the simulations with \sigis = 6 \kms, no sightline had more than one component meaning we are not able to calculate \sigls for any of these simulations.

Figure \ref{fig:res1D} shows the measured \bturbs (top panels) and \sigls (bottom panels) for each simulation as a function of input \sigi. Each column corresponds to a different density box, as labeled at the top. In each panel, we show the 1$-$to$-$1 relation as a black dashed line.
Each simulation produced at least 5 total absorbers of each ion, with as many as 23, 16, 18, and 22 being found in a single box for \SiII, \SiIV, \CIV, and \OVI, respectively. The simulations with larger \sigis injected produced more components than those will smaller velocities. The sightlines used to derive the \sigls values contained 2 to 4 absorbers, on average. While a single sightline had 8 absorbers, the rest had 6 or fewer.

All components were found to have \bturbs $< 40$ \kms, with the vast majority ($\approx$85\%) being narrower than 20 \kms.
In addition, as \sigis increases, the median \bturbs remain relatively flat. 
This can be seen in the best-fit lines, which were
\begin{eqnarray}
b_\mathrm{turb} = &(0.135 \pm 0.003) \sigma_\mathrm{1D} + (7.69 \pm 4.23), & \\
b_\mathrm{turb} = &(0.067 \pm 0.005) \sigma_\mathrm{1D}+ (6.85 \pm 6.42), & \\
b_\mathrm{turb} = &(0.013 \pm 0.006) \sigma_\mathrm{1D}+ (14.49 \pm 7.37), &
\end{eqnarray}
for $nL_{\rm drive}= 10^{21}$, $10^{20},$ and $10^{19}$ cm$^{-2},$ respectively.
Performing Kendall $\uptau$ tests of these samples gave coefficients of 0.467, 0.333, and 0.200. 
These suggest that \bturbs may not be able to accurately infer the amount of turbulence that is present on small scales.

This is likely because this quantity measures turbulence on the scale of individual warm clouds within the larger CGM, which do not sample the overall random motions that exist on larger scales. Moreover, as the dissipation of turbulence leads to heating, the level of turbulence within a given cloud is likely to be correlated with the cloud temperature, and hence with the presence of the ions we are using to measure it. Our results imply that these effects are strong enough to make \bturbs much more dependent on the typical conditions that support the existence of \CIVs and \SiIVs than on the global turbulent properties of the medium.

On the other hand, \sigls does a much better job tracing \sigi. In this case
\begin{eqnarray}
\sigma_\mathrm{LOS} = &(1.005 \pm 0.002) \sigma_\mathrm{1D} + (9.68 \pm 2.94), & \\
\sigma_\mathrm{LOS} = &(0.501 \pm 0.091) \sigma_\mathrm{1D}+ (4.23 \pm 139.80), & \\
\sigma_\mathrm{LOS} = &(0.336 \pm 0.092) \sigma_\mathrm{1D}+ (14.69 \pm 169.60), &
\end{eqnarray}
for $nL_{\rm drive}= 10^{21}$, $10^{20}$, and $10^{19}$ cm$^{-2}$, respectively, and Kendall $\uptau$ tests of these samples gave coefficients of 1.00, 0.8, and 0.34, respectively. This means that although the random motions within individual warm clouds are not well correlated with the overall turbulent velocity, the level of random motions {\em between} pairs of warm clouds is strongly correlated with overall turbulence. This is likely because the relative velocities of clouds are able to sample random motions on scales much larger than the clouds themselves, providing a good estimate of the overall level of turbulence in the system.

Looking at each gas phase independently shows that the best tracer of \sigis depends on the density of the gas. \sigl(\OVI) followed \sigis in the $10^{19}$ cm$^{-2}$ density box while \sigl(\SiII) produced an anti-correlation with increasing \sigi. In the $10^{21}$ cm$^{-2}$ density box, however, \sigl(\SiII) was able to trace \sigis just as well as \sigl(\CIV) and \sigl(\SiIV). Meanwhile, \sigl(\CIV) and \sigl(\SiIV) did equally well in the $10^{20}$ cm$^{-2}$ density simulation.

To determine if these results are robust to viewing angle, we generated additional sightlines through these simulations along all three directions. The \bturbs and \sigls values were consistent regardless of the axis probed, which is expected for these homogeneous simulations.

\begin{deluxetable*}{CCCCCCC}[t]
    \label{tab:large_results}
    \centering
    \tablecaption{Mean turbulence results through the large-scale simulations.}
    \tablehead{
    \colhead{$\rho$} & \colhead{\bturb} & \colhead{$b_\mathrm{therm}$} & \colhead{\sigls (\SiII)} & \colhead{\sigls (\SiIV)} & \colhead{\sigls (\CIV)} & \colhead{\sigls (\OVI)}\\
    \colhead{$\left[\mathrm{kpc}\right]$} & \colhead{$\left[\mathrm{km\ s}^{-1}\right]$} & \colhead{$\left[\mathrm{km\ s}^{-1}\right]$} & \colhead{$\left[\mathrm{km\ s}^{-1}\right]$} & \colhead{$\left[\mathrm{km\ s}^{-1}\right]$} & \colhead{$\left[\mathrm{km\ s}^{-1}\right]$} & \colhead{$\left[\mathrm{km\ s}^{-1}\right]$}}
    \startdata
    \multicolumn{7}{c}{MHD Simulation}\\
    \hline
    0 & 34.3 \pm 19.6 & 16.8 \pm 14.1 & 100.2 & 67.9 & 74.2 & 39.2\\ 
    10 & 15.3 \pm 8.4 & 23.4 \pm 8.2 & 78.0 \pm 21.3 & 72.2 \pm 21.9 & 73.7 \pm 16.3 & 77.0 \pm 14.4\\ 
    14 & 14.5 \pm 9.5 & 16.5 \pm 7.1 & 64.9 \pm 13.5 & 77.1 \pm 21.3 & 81.4 \pm 13.6 & 87.9 \pm 17.3\\ 
    25 & 18.3 \pm 12.4 & 24.0 \pm 23.2 & 71.0 \pm 11.0 & 78.7 \pm 39.8 & 58.9 \pm 19.8 & 66.5 \pm 19.3\\ 
    35 & 15.7 \pm 8.7 & 16.1 \pm 9.6 & 49.9 \pm 27.4 & 57.9 \pm 27.9 & 58.8 \pm 26.9 & 50.7 \pm 21.2\\ 
    50 & 11.3 \pm 4.8 & 17.2 \pm 8.3 & 38.0 \pm 2.3 & 60.7 \pm 8.6 & 55.2 \pm 3.8 & 38.7 \pm 20.2\\ 
    71 & 8.1 \pm 3.2 & 16.8 \pm 4.7 & 49.1 \pm 19.8 & 54.6 \pm 9.0 & 39.6 \pm 18.3 & 58.8 \pm 17.0\\ 
    75 $-$ 110 & 12.0 \pm 5.0 & 14.2 \pm 7.1 & 32.7 \pm 23.5 & 35.1 \pm 26.8 & 35.0 \pm 28.7 & 35.1 \pm 22.9\\ 
    \hline
    \multicolumn{7}{c}{Hydro Simulation} \\
    \hline
    0 & 16.9 \pm 11.2 & 17.8 \pm 11.4 & 127.5 & 132.9 & 131.0 & 152.2\\ 
    10 & 11.2 \pm 5.7 & 13.4 \pm 8.7 & 70.3 \pm 16.8 & 84.8 \pm 15.1 & 83.1 \pm 18.6 & 100.9 \pm 7.5\\ 
    14 & 10.1 \pm 2.1 & 15.8 \pm 8.6 & 73.4 \pm 29.4 & 109.0 \pm 32.1 & 107.6 \pm 32.1 & 123.1 \pm 10.0\\ 
    25 & 10.6 \pm 5.6 & 12.4 \pm 6.5 & 85.0 \pm 26.0 & 82.7 \pm 16.9 & 82.7 \pm 17.4 & 80.4 \pm 27.8\\ 
    35 & 9.9 \pm 3.6 & 13.1 \pm 6.3 & 75.6 \pm 20.2 & 79.5 \pm 19.8 & 82.1 \pm 17.6 & 87.0 \pm 20.1\\ 
    50 & 12.6 \pm 4.9 & 17.9 \pm 7.0 & 53.0 \pm 30.7 & 56.8 \pm 21.2 & 59.2 \pm 19.8 & 69.8 \pm 22.7\\ 
    71 & 12.0 \pm 7.9 & 14.3 \pm 6.5 & 25.3 \pm 20.0 & 29.9 \pm 16.0 & 28.2 \pm 15.0 & 24.7 \pm 9.1\\ 
    75 $-$ 110 & 9.4 \pm 3.8 & 19.2 \pm 13.3 & 33.1 \pm 17.2 & 38.8 \pm 16.3 & 41.6 \pm 14.0 & 43.0 \pm 25.1\\ 
    \enddata
    \tablenotetext{}{The upper results correspond to the MHD simulation while the lower results are for the Hyrdo simulation with the uncertainties showing the standard deviation on the distribution.The uncertainties show the standard deviation on the distribution. At the simulation center, there is only one possible sightline and so are unable to report a distribution for the \sigls results.}
\end{deluxetable*}

\subsection{Large-Scale} \label{sect:large}

While the small-scale simulations tell us about how turbulence impacts individual CGM regions, the large-scale simulations allow us to investigate the more global properties of CGM turbulence. Our sightlines probe these simulations at impact parameters within 110 kpc of their centers ($\rho$ / R$_\mathrm{vir} \leq 0.5$). Beyond this limit, \CIVs and \SiIVs were not abundant enough to measure \bturbs robustly and so they are not included in our analysis. The radial extent of our sightlines is comparable to what most absorption studies of the CGM have access to \citep{Tumlinson2013,Bordoloi2014,Borthakur2015}, although occasionally observations are available at or beyond R$_\mathrm{vir}$.

In Table \ref{tab:large_results} we report the average and spread of the turbulent measures in each of our simulations as a function of impact parameter. Here we bin the sightlines with $75 \leq \rho < 110$ kpc so that a similar number of sightlines and components were used to calculate the values in each row. At impact parameters of $\rho\ <\ 50$ kpc, the Hydro box produced nearly 2 more components per sightline than the MHD simulation, while both runs produced a similar number of components at larger radii. This difference is likely due to the fact that magnetic pressure causes the colder regions to be more puffed up in the MHD simulation, as seen in Figure \ref{fig:large_temp_slice}, and this leads to lower overall columns. As discussed in \citep{Buie2022} in the MHD simulation such dynamically important fields arise only within $\approx$ 50 kpc of the center of the halo (see Figure 10 in that work), explaining why the differences we see here are confined to $\rho \lesssim 50$ kpc.
Sightlines through the Hydro simulation produced 124 \SiIIs absorbers, 155 \SiIVs absorbers, 165 \CIVs absorbers, and 163 \OVIs absorbers while the MHD simulation produced 110, 138, 142, and 122, respectively.

\begin{figure*}
    \centering
    \includegraphics[width=\linewidth]{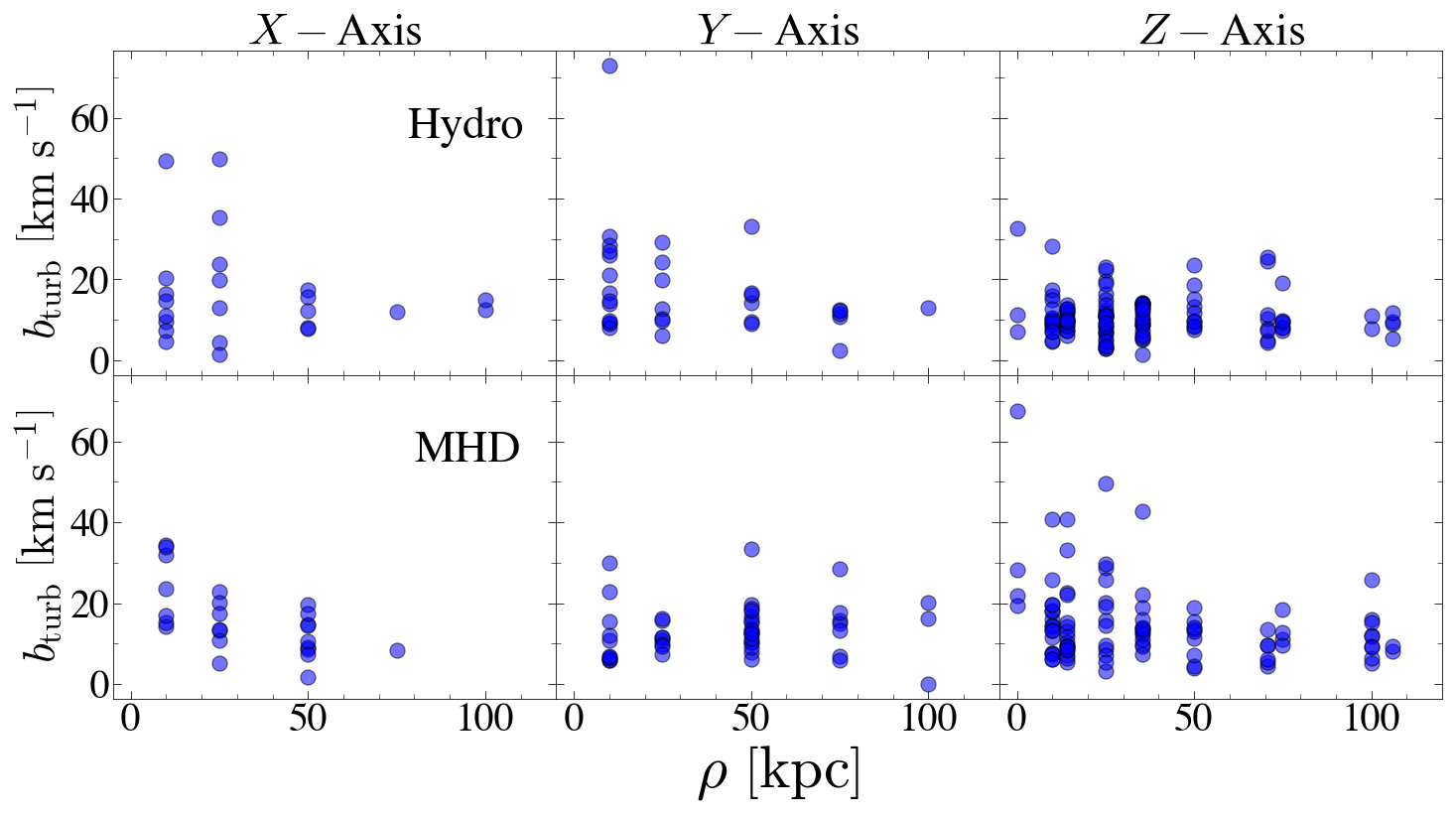}
    \caption{Component turbulent measure, \bturb, in the large-scale simulations as a function of distance from the center of the simulation, $\rho$. Results through the Hydro box are shown on the top panel while the MHD results are on the bottom. Each column corresponds to a different axis being probed by the sightlines, where $y$ is along the axis of rotation, and $x$ and $z$ are perpendicular to it.  Sightlines along the $x-$ axis and $y-$axes are a subset of those along the $z-$ axis and so fewer points are present in these panels.}
    \label{fig:large_bturb}
\end{figure*}

The right column of Figure \ref{fig:large_bturb} shows the \bturbs measured for each large-scale simulation along the $z-$axis} as a function of $\rho$, with the Hydro box in the top panel and the MHD box in the bottom panel. Both simulations produced clouds with \bturbs $\leq 20$ \kmss at all $\rho$, and several components with large \bturbs were found at $\rho \lesssim 35$ kpc, i.e., where the amount of injected turbulence is highest. In the MHD simulation, 5 components had \bturbs $\geq 40$ \kmss while none were found in the small-scale or Hydro simulations with such large turbulent velocities. In fact, the Hydro simulation nearly always produced components with smaller \bturbs velocities than the MHD simulation. Again this difference is likely related to the fact that the colder regions are more puffed up in this simulation, and thus they experience the larger turbulent velocities that arise at larger physical scales. 
This is consistent with the fact that the distribution of \bturbs values in the two runs become similar at large radii where magnetic effects are negligible, leading to a slight anti-correlation between \bturbs and $\rho$ in the MHD run.

Our large-scale MHD simulation also produced a similar fraction of sightlines with \bturbs $= 20 - 40$ \kmss as the small-scale simulations ($\approx$13\% compared to $\approx$16\% in the small-scale simulations). In both large-scale simulations, the distribution of \bturbs is dominated by the components with the smallest velocities $\approx$82\% as compared to $\approx$85\% for the small-scale simulations. From these results, it is clear that \bturbs does not provide a good tracer of the true turbulent velocities. 

\begin{figure*}
    \centering
    \includegraphics[width=\linewidth]{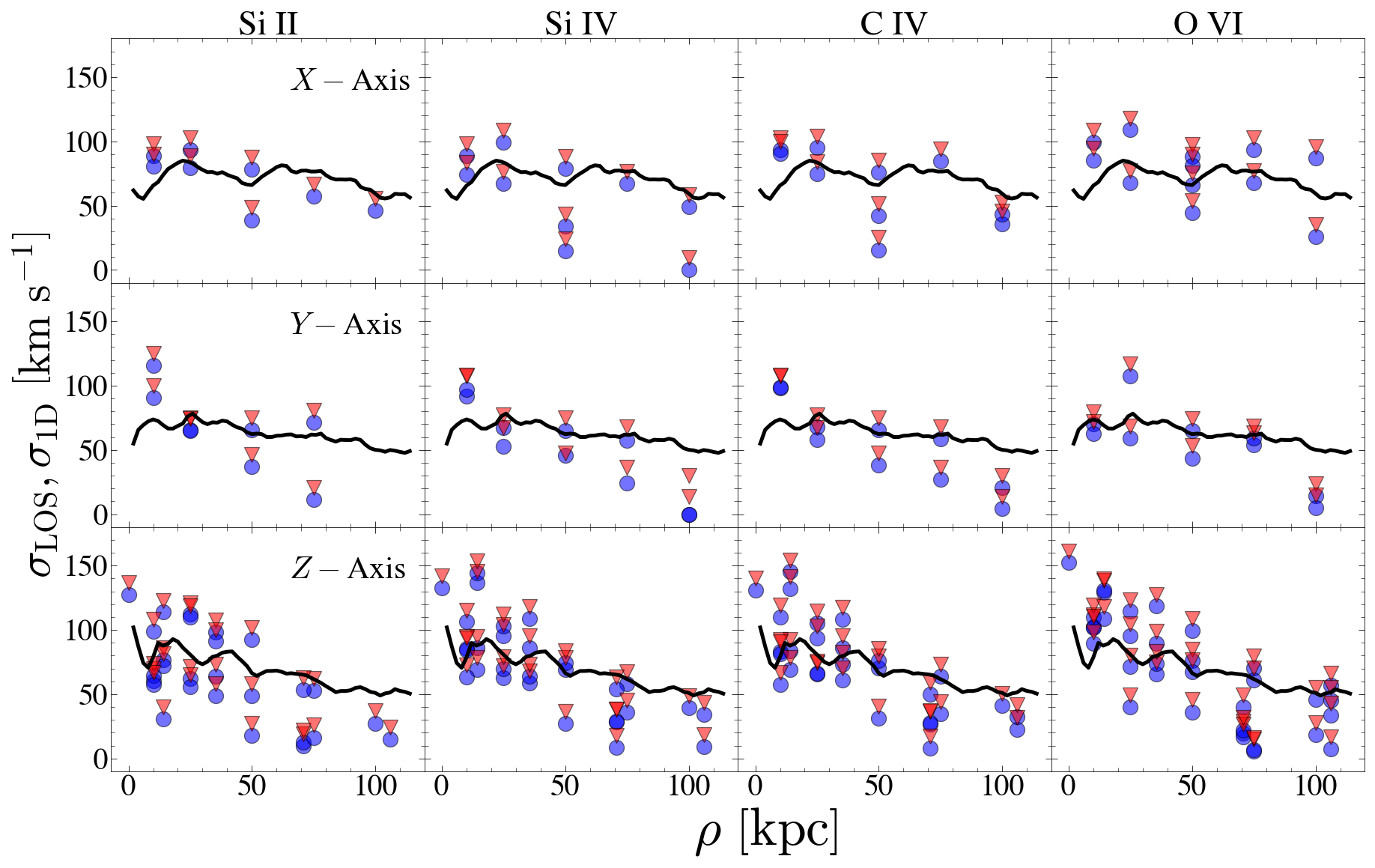}
    \caption{Sightline turbulent measure, \sigl, in the Hydro simulation as a function of distance from the center of the simulation, $\rho$. Our measured values are shown in as blue circles while the red triangles shows the measurements after applying the line of best fit from the $nL_\mathrm{drive} = 10^{21}$ cm$^{-2}$ small-scale simulation. The errors on \sigls were smaller than the point sizes and so are not shown. Each column corresponds to a different ion, which is labeled at the top while each row corresponds to a different axis being probed by the sightlines. The ``true'' \sigls is shown in each panel as a black line as defined in the text.}
    \label{fig:Hydro_sigma_LOS}
\end{figure*}

\begin{figure*}
    \centering
    \includegraphics[width=\linewidth]{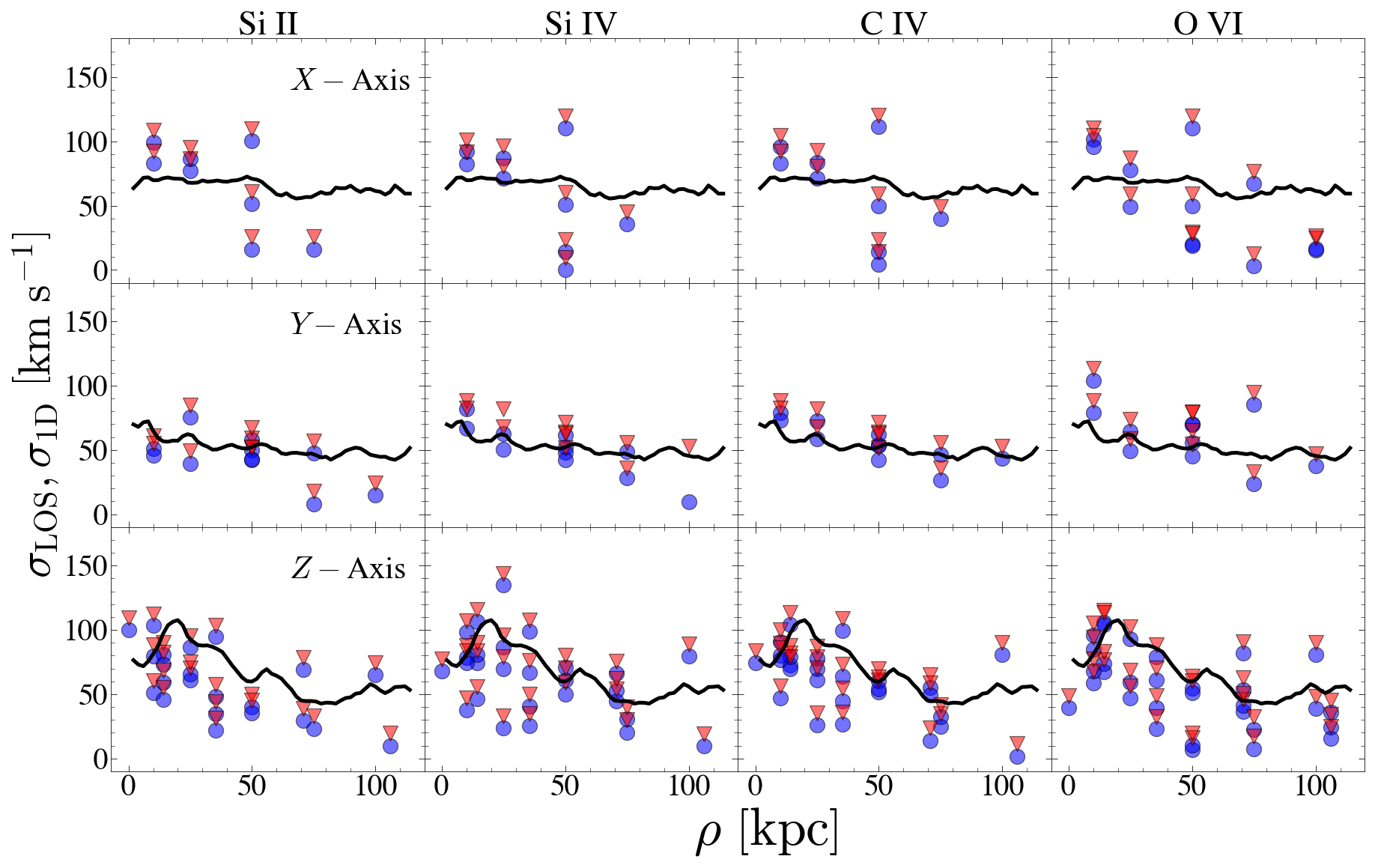}
    \caption{Same as Figure~\ref{fig:Hydro_sigma_LOS} except for the measurements through the MHD simulation.}
    \label{fig:MHD_sigma_LOS}
\end{figure*}

The bottom row of Figures~\ref{fig:Hydro_sigma_LOS} and \ref{fig:MHD_sigma_LOS} shows our measurements of \sigls from the Hydro and MHD simulations, respectively, as a function of $\rho$. 
Here each column shows the results of a different ion, labeled in order of increasing ionization potential. To determine the amount of turbulence as a function of radius, we directly measured the volume average, or ``true'', \sigls using all the cells which have temperatures $<$10$^5$ K and at the same projected radius from the center of the simulations, generating straws in the axis being probed. For each radius, we computed the volume average radial velocity and the volume average rotational velocity. 
We then subtracted these from each cell to leave only the random component of the line-of-sight velocity, which we then used to compute $\sigma_{\rm 1D}$. This measure is shown in the figures as solid black lines, which we compare both to our direct measurement of \sigls (blue points), as well as \sigis estimated from the best fit from the $nL_\mathrm{drive} = 10^{21}$ cm$^{-2}$ small-scale simulations ($\sigma_{\rm 1D} \approx \sigma_{\rm LOS} + 10$ \kms; red triangles).
The sightlines used to derive the \sigls values shown here contained 2 to 5 absorbers on average, with some sightlines having as many as 9 individual absorbers.

A clear radial trend is seen in each panel, with sightlines closer to the simulation centers having larger \sigls values than those further away, and with the Hydro sightlines tending to have larger velocities overall. This decrease appears to flatten out around 75 kpc, though the spread is still rather large. In both simulations, many of the \sigls points are found below the true $\sigma_{\rm 1D}$ line, though they do follow the same trend with radius. Applying the line of best fit from the small-scale simulation causes some points in each panel to become roughly consistent with the true \sigls line in most panels, providing a general estimate of the underlying turbulence, especially in the MHD run. These findings suggest that the actual large-scale turbulent velocities in the CGM are not measured well on a case-by-case basis by \sigl, but a simple estimate of $\sigma_{\rm 1D} \approx \sigma_{\rm LOS} + 10$ \kmss provides a reasonable picture of the underlying turbulence on an average basis.

Unlike in the small-scale simulations, we do not see much difference between the gas phases in either the MHD or Hydro simulations. The radial trends are largely the same in each ion, as shown in Figures~\ref{fig:Hydro_sigma_LOS} and \ref{fig:MHD_sigma_LOS}.

Similar to the small-scale simulations, we generate additional sightlines through the large-scale simulations along their $y-$axes (i.e., their axes of rotation) and $x-$axis to determine whether our results are robust to viewing angle relative to the simulated galaxy. The \bturbs values from these sightlines are shown in the center and left columns of Figure~\ref{fig:large_bturb}, respectively. These show a slight trend with radius, consistent with what we found along the $z-$axes. The center and top rows of Figure~\ref{fig:Hydro_sigma_LOS} and \ref{fig:MHD_sigma_LOS} show the \sigls values from these sightlines. These produced a similar radial decline as was seen in our previous results. These indicates that our results are robust to the viewing angle of the sightline.

\subsection{Comparison to Observational CGM Surveys}

Finally, we compare our large-scale simulations with observational surveys. 
The \bturbs measure was employed by \citet{Rudie2019} to constrain the turbulent velocities for 8 galaxies at $z \approx 2$ with stellar masses of M$_\star \approx 10^{9} - 10^{11}$ \Ms and a median halo mass of M$_\mathrm{halo} = 10^{11.9}$ \M. We find a similar distribution in our MAIHEM simulations, with most components having \bturbs $\leq 15$ \kmss and a few being as high as 75 \kms. While they did not find an anti-correlation with $\rho$, they attribute this to the small sample size, stating that the complete sample does show this trend. 

\citet{Werk2016} decomposed the Doppler $b$ parameters of their \OVIs absorbers in the COS-Halos sample into thermal and non-thermal components, as we have done with \CIVs and \SiIV.
However, they assume a range of thermal contributions (6.4 \kmss $\lesssim b_\mathrm{therm}$(\OVI) $\lesssim 16.2$ \kms) rather than the method we use, described in \S \ref{sect:def_turb}.
With this, they found an average turbulent velocity range of $\approx$40 $-$ 50 \kms. While this is higher than the \bturbs values derived here from the Doppler widths of \CIVs and \SiIV, these are in line with our \sigl(\OVI) velocities in Figures \ref{fig:Hydro_sigma_LOS} and \ref{fig:MHD_sigma_LOS}.

\begin{figure}
    \centering
    \includegraphics[width=\linewidth]{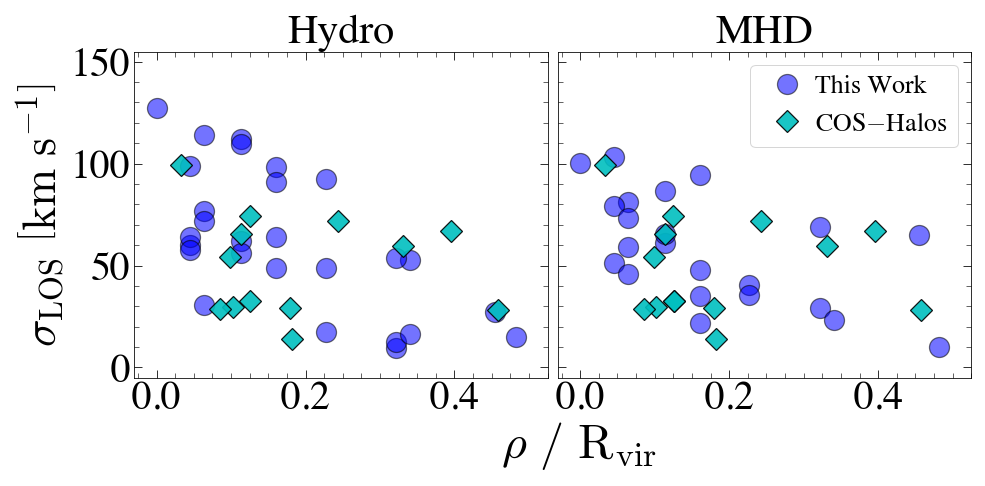}
    \caption{Comparing \sigl(\SiII) measured through the large-scale simulations to the COS-Halos observational survey as a function of impact parameter normalized by the virial radius, $\rho\ /\ \mathrm{R}_\mathrm{vir}$. Blue circles show our measurements through the Hydro and MHD simulations on the left and right panels, respectively while values derived from COS-Halos measurements \citep{Werk2013} are shown as cyan diamonds.}
    \label{fig:sigma_comp}
\end{figure}

Using the velocity centroids of \SiIIs absorbers from the COS-Halos survey \citep{Werk2013} allows us to compare our \sigls values to observations, which we show in Figure \ref{fig:sigma_comp}. Both samples show a decrease in \sigl(\SiII) with increasing $\rho$ / R$_\mathrm{vir}$, with no absorbers being found beyond $\rho~\approx~0.5$~R$_\mathrm{vir}$. Similar velocities were found in each distribution, suggesting the turbulence in our large-scale MAIHEM simulations is comparable to that in the CGM of COS-Halos galaxies.

\section{Conclusions} \label{sect:summary}

The properties of the CGM are essential to determining the history of galaxy evolution, but our information about this medium is mostly limited to information inferred from QSO absorption-line measurements. Such measurements uncover a multiphase medium that strongly implies the presence of significant turbulence, yet the properties of this turbulence remain largely unknown. Here we have analyzed a suite of CGM simulations to determine how well this underlying turbulence relates to two commonly-observed quantities: the non-thermal width of absorption-line components (\bturb) and the line-of-sight velocity dispersion between components (\sigl). Our simulations made use of the MAIHEM code to track non-equilibrium ionization effects, and they consisted both of small-patch and global simulations. This allowed us to probe turbulent motion on both large and small scales, as well as determine how magnetic fields impact \sigls and \bturb.

Our analysis of both types of simulations showed that the turbulence in individual warm absorbers, as measured by \bturb, did not trace the turbulent velocities in the media. In our small-scale simulations, the majority of absorbers were narrower than 20 \kmss even in cases in which the overall one-dimensional turbulent velocities were as large as 58 \kms. Furthermore, the mean \bturbs did not change with increasing \sigi, though the spread tended to get larger. Kendall $\uptau$ tests show that \bturbs and \sigis are not well correlated, likely because \bturbs measures random motions within individual warm clouds, which do not sample the overall CGM motions. The measure also did a poor job at tracing the turbulent field in the large-scale Hydro simulations, although it was able to show an anti-correlation with radius in the large-scale MHD case, indicative of the declining importance of magnetic pressure at large radii in this simulation.

On the other hand, the line-of-sight velocity dispersion between components, \sigl, was able to trace the injected turbulence at all scales. In the small-scale simulation with $nL_\mathrm{drive} = 10^{21}$ cm$^{-2}$, we found that \sigis $\approx$ \sigls $+ 10$ \kmss provided a good description of our results. This trend was shown to be the strongest observed in our study, with a Kendall coefficient of 1. Correlations were also seen in the lower density boxes; however, the trends were somewhat less strong, with \sigis depending more steeply on \sigl, such that true turbulent velocity dispersion was often underestimated in the less turbulent boxes. On these scales, we find evidence for the trends depending weakly on the gas phase probing \sigis and are robust to viewing angle.

The large-scale simulations, both with and without MHD physics, produced an anti-correlation between $\rho$ and \sigls that flattened out at $\rho \approx 75$ kpc. In these boxes, the measured \sigls was consistently below the injected turbulence at all $\rho$, with few exceptions, though \sigls and \sigis followed the same radial trend. Unlike in the small-scale simulations, this did not depend on the gas phase being considered. Applying the line of best fit from the small-scale simulation causes some points in each panel to become roughly consistent with the true \sigis line, providing a general estimate of the underlying turbulence, especially in the case of the MHD run. These findings suggest that the actual large-scale turbulent velocities in the CGM are not well measured on a case-by-case basis by \sigl, but a simple of estimate of \sigis $\approx$ \sigls $+ 10$ \kmss provides a reasonable picture of the underlying turbulence on an average basis regardless of viewing angle.

The results from our global simulations are also consistent with those found in recent observations, which indicate that most individual absorbers have \bturbs $\leq 15$ \kms, and that the line-of-sight velocity dispersion between absorbers have $10 \, {\rm km\ s}^{-1} \lesssim \sigma_{\rm LOS} \lesssim 150$ \kms, with \sigls declining strongly with impact parameter. This agreement suggests that the turbulence in our large-scale MAIHEM simulations is comparable to that in the CGM probed by these samples, and that strong correlation between $\sigma_{\rm 1D}$ and $\sigma_{\rm LOS}$ observed in our study may provide to be useful in analyzing observed systems. Future observational and theoretical studies are needed to help refine our understanding of this connection and its applicability to constraining circumgalactic turbulence with QSO absorption-line measurements.

We would like to acknowledge Sanchayeeta Borthakur and Tyler McCabe for their helpful discussions that greatly improved this manuscript. The simulations used in this work were run on the NASA Pleiades supercomputer maintained by the Science Mission Directorate HighEnd Computing program and the Stampede2 supercomputer at the Texas Advanced Computing Center (TACC) through Extreme Science and Engineering Discovery Environment (XSEDE) resources under grants TGAST130021 and TGPHY200071. E.S. acknowledges support from NASA grant 80NSSC22K1265. 
The authors acknowledge the native people and the land that Arizona State University’s campuses are located in the Salt River Valley. The ancestral territories of Indigenous peoples, including the Akimel O’odham (Pima) and Pee Posh (Maricopa) Indian Communities, whose care and keeping of these lands allows us to be here today.

\software{FLASH (\citealt{Fryxell2000}; v4.5), matplotlib (\citealt{Hunter2007}; 3.2.2), yt \citep{Turk2011}, numpy (\citealt{Harris2020}; v1.22.0), pandas (\citealt{Reback2021}; 1.3.5)}

\bibliographystyle{aasjournal}
\bibliography{References}

\begin{thebibliography}{}
\expandafter\ifx\csname natexlab\endcsname\relax\def\natexlab#1{#1}\fi
\providecommand{\url}[1]{\href{#1}{#1}}
\providecommand{\dodoi}[1]{doi:~\href{http://doi.org/#1}{\nolinkurl{#1}}}
\providecommand{\doeprint}[1]{\href{http://ascl.net/#1}{\nolinkurl{http://ascl.net/#1}}}
\providecommand{\doarXiv}[1]{\href{https://arxiv.org/abs/#1}{\nolinkurl{https://arxiv.org/abs/#1}}}

\bibitem[{{Arav} {et~al.}(2013){Arav}, {Borguet}, {Chamberlain}, {Edmonds}, \&
  {Danforth}}]{Arav2013}
{Arav}, N., {Borguet}, B., {Chamberlain}, C., {Edmonds}, D., \& {Danforth}, C.
  2013, \mnras, 436, 3286, \dodoi{10.1093/mnras/stt1812}

\bibitem[{{Bordoloi} {et~al.}(2014){Bordoloi}, {Tumlinson}, {Werk},
  {Oppenheimer}, {Peeples}, {Prochaska}, {Tripp}, {Katz}, {Dav{\'e}}, {Fox},
  {Thom}, {Ford}, {Weinberg}, {Burchett}, \& {Kollmeier}}]{Bordoloi2014}
{Bordoloi}, R., {Tumlinson}, J., {Werk}, J.~K., {et~al.} 2014, \apj, 796, 136,
  \dodoi{10.1088/0004-637X/796/2/136}

\bibitem[{{Borthakur} {et~al.}(2015){Borthakur}, {Heckman}, {Tumlinson},
  {Bordoloi}, {Thom}, {Catinella}, {Schiminovich}, {Dav{\'e}}, {Kauffmann},
  {Moran}, \& {Saintonge}}]{Borthakur2015}
{Borthakur}, S., {Heckman}, T., {Tumlinson}, J., {et~al.} 2015, \apj, 813, 46,
  \dodoi{10.1088/0004-637X/813/1/46}

\bibitem[{{Borthakur} {et~al.}(2016){Borthakur}, {Heckman}, {Tumlinson},
  {Bordoloi}, {Kauffmann}, {Catinella}, {Schiminovich}, {Dav{\'e}}, {Moran}, \&
  {Saintonge}}]{Borthakur2016}
---. 2016, \apj, 833, 259, \dodoi{10.3847/1538-4357/833/2/259}

\bibitem[{{Buie} {et~al.}(2020{\natexlab{a}}){Buie}, {Fumagalli}, \&
  {Scannapieco}}]{Buie2020a}
{Buie}, Edward, I., {Fumagalli}, M., \& {Scannapieco}, E. 2020{\natexlab{a}},
  \apj, 890, 33, \dodoi{10.3847/1538-4357/ab65bc}

\bibitem[{{Buie} {et~al.}(2018){Buie}, {Gray}, \& {Scannapieco}}]{Buie2018}
{Buie}, Edward, I., {Gray}, W.~J., \& {Scannapieco}, E. 2018, \apj, 864, 114,
  \dodoi{10.3847/1538-4357/aad8bd}

\bibitem[{{Buie} {et~al.}(2020{\natexlab{b}}){Buie}, {Gray}, {Scannapieco}, \&
  {Safarzadeh}}]{Buie2020b}
{Buie}, Edward, I., {Gray}, W.~J., {Scannapieco}, E., \& {Safarzadeh}, M.
  2020{\natexlab{b}}, \apj, 896, 136, \dodoi{10.3847/1538-4357/ab9535}

\bibitem[{{Buie} {et~al.}(2022){Buie}, {Scannapieco}, \& {Mark
  Voit}}]{Buie2022}
{Buie}, Edward, I., {Scannapieco}, E., \& {Mark Voit}, G. 2022, \apj, 927, 30,
  \dodoi{10.3847/1538-4357/ac4bc2}

\bibitem[{{Burchett} {et~al.}(2015){Burchett}, {Tripp}, {Prochaska}, {Werk},
  {Tumlinson}, {O'Meara}, {Bordoloi}, {Katz}, \& {Willmer}}]{Burchett2015}
{Burchett}, J.~N., {Tripp}, T.~M., {Prochaska}, J.~X., {et~al.} 2015, \apj,
  815, 91, \dodoi{10.1088/0004-637X/815/2/91}

\bibitem[{Burchett {et~al.}(2019)Burchett, Tripp, Prochaska, Werk, Tumlinson,
  Howk, Willmer, Lehner, Meiring, Bowen, Bordoloi, Peeples, Jenkins, O'Meara,
  Tejos, \& Katz}]{Burchett2019}
Burchett, J.~N., Tripp, T.~M., Prochaska, J.~X., {et~al.} 2019, \apjl, 877,
  L20, \dodoi{10.3847/2041-8213/ab1f7f}

\bibitem[{{Chen} {et~al.}(2020){Chen}, {Zahedy}, {Boettcher}, {Cooper},
  {Johnson}, {Rudie}, {Chen}, {Walth}, {Cantalupo}, {Cooksey},
  {Faucher-Gigu{\`e}re}, {Greene}, {Lopez}, {Mulchaey}, {Penton}, {Petitjean},
  {Putman}, {Rafelski}, {Rauch}, {Schaye}, {Simcoe}, \& {Weiner}}]{Chen2020}
{Chen}, H.-W., {Zahedy}, F.~S., {Boettcher}, E., {et~al.} 2020, \mnras, 497,
  498, \dodoi{10.1093/mnras/staa1773}

\bibitem[{{Chisholm} {et~al.}(2017){Chisholm}, {Tremonti}, {Leitherer}, \&
  {Chen}}]{Chisholm2017}
{Chisholm}, J., {Tremonti}, C.~A., {Leitherer}, C., \& {Chen}, Y. 2017, \mnras,
  469, 4831, \dodoi{10.1093/mnras/stx1164}

\bibitem[{{Comparat} {et~al.}(2022){Comparat}, {Truong}, {Merloni},
  {Pillepich}, {Ponti}, {Driver}, {Bellstedt}, {Liske}, {Aird}, {Br{\"u}ggen},
  {Bulbul}, {Davies}, {Villalba}, {Georgakakis}, {Haberl}, {Liu}, {Maitra},
  {Nandra}, {Popesso}, {Predehl}, {Robotham}, {Salvato}, {Thorne}, \&
  {Zhang}}]{Comparat2022}
{Comparat}, J., {Truong}, N., {Merloni}, A., {et~al.} 2022, \aap, 666, A156,
  \dodoi{10.1051/0004-6361/202243101}

\bibitem[{{Croton} {et~al.}(2006){Croton}, {Springel}, {White}, {De Lucia},
  {Frenk}, {Gao}, {Jenkins}, {Kauffmann}, {Navarro}, \& {Yoshida}}]{Croton2006}
{Croton}, D.~J., {Springel}, V., {White}, S. D.~M., {et~al.} 2006, \mnras, 365,
  11, \dodoi{10.1111/j.1365-2966.2005.09675.x10.48550/arXiv.astro-ph/0508046}

\bibitem[{{Dekel} \& {Birnboim}(2006)}]{Dekel2006}
{Dekel}, A., \& {Birnboim}, Y. 2006, \mnras, 368, 2

\bibitem[{{Di Matteo} {et~al.}(2005){Di Matteo}, {Springel}, \&
  {Hernquist}}]{DiMatteo2005}
{Di Matteo}, T., {Springel}, V., \& {Hernquist}, L. 2005, \nat, 433, 604,
  \dodoi{10.1038/nature0333510.48550/arXiv.astro-ph/0502199}

\bibitem[{{Eckert} {et~al.}(2019){Eckert}, {Ghirardini}, {Ettori}, {Rasia},
  {Biffi}, {Pointecouteau}, {Rossetti}, {Molendi}, {Vazza}, {Gastaldello},
  {Gaspari}, {De Grandi}, {Ghizzardi}, {Bourdin}, {Tchernin}, \&
  {Roncarelli}}]{Eckert2019}
{Eckert}, D., {Ghirardini}, V., {Ettori}, S., {et~al.} 2019, \aap, 621, A40,
  \dodoi{10.1051/0004-6361/201833324}

\bibitem[{{Erb} {et~al.}(2022){Erb}, {Li}, {Steidel}, {Chen}, {Gronke},
  {Strom}, {Trainor}, \& {Rudie}}]{Erb2022}
{Erb}, D.~K., {Li}, Z., {Steidel}, C.~C., {et~al.} 2022, arXiv e-prints,
  arXiv:2210.02465, \dodoi{10.48550/arXiv.2210.02465}

\bibitem[{{Fabian}(2012)}]{Fabian2012}
{Fabian}, A.~C. 2012, \araa, 50, 455,
  \dodoi{10.1146/annurev-astro-081811-125521}

\bibitem[{{Federrath} {et~al.}(2010){Federrath}, {Roman-Duval}, {Klessen},
  {Schmidt}, \& {Mac Low}}]{Federrath2010}
{Federrath}, C., {Roman-Duval}, J., {Klessen}, R.~S., {Schmidt}, W., \& {Mac
  Low}, M.~M. 2010, \aap, 512, A81, \dodoi{10.1051/0004-6361/200912437}

\bibitem[{{Fielding} {et~al.}(2017){Fielding}, {Quataert}, {McCourt}, \&
  {Thompson}}]{Fielding2007}
{Fielding}, D., {Quataert}, E., {McCourt}, M., \& {Thompson}, T.~A. 2017,
  \mnras, 466, 3810, \dodoi{10.1093/mnras/stw3326}

\bibitem[{{Fryxell} {et~al.}(2000){Fryxell}, {Olson}, {Ricker}, {Timmes},
  {Zingale}, {Lamb}, {MacNeice}, {Rosner}, {Truran}, \& {Tufo}}]{Fryxell2000}
{Fryxell}, B., {Olson}, K., {Ricker}, P., {et~al.} 2000, \apjs, 131, 273,
  \dodoi{10.1086/317361}

\bibitem[{{Gray} \& {Scannapieco}(2016)}]{Gray2016}
{Gray}, W.~J., \& {Scannapieco}, E. 2016, \apj, 818, 198,
  \dodoi{10.3847/0004-637X/818/2/198}

\bibitem[{{Gray} \& {Scannapieco}(2017)}]{Gray2017}
---. 2017, \apj, 849, 132, \dodoi{10.3847/1538-4357/aa9121}

\bibitem[{{Gray} {et~al.}(2015){Gray}, {Scannapieco}, \& {Kasen}}]{Gray2015}
{Gray}, W.~J., {Scannapieco}, E., \& {Kasen}, D. 2015, \apj, 801, 107,
  \dodoi{10.1088/0004-637X/801/2/107}

\bibitem[{{Haardt} \& {Madau}(2012)}]{HM2012}
{Haardt}, F., \& {Madau}, P. 2012, \apj, 746, 125,
  \dodoi{10.1088/0004-637X/746/2/125}

\bibitem[{{Hamann} {et~al.}(2011){Hamann}, {Kanekar}, {Prochaska}, {Murphy},
  {Ellison}, {Malec}, {Milutinovic}, \& {Ubachs}}]{Hamann2011}
{Hamann}, F., {Kanekar}, N., {Prochaska}, J.~X., {et~al.} 2011, \mnras, 410,
  1957, \dodoi{10.1111/j.1365-2966.2010.17575.x}

\bibitem[{{Harris} {et~al.}(2020){Harris}, {Millman}, {van der Walt},
  {Gommers}, {Virtanen}, {Cournapeau}, {Wieser}, {Taylor}, {Berg}, {Smith},
  {Kern}, {Picus}, {Hoyer}, {van Kerkwijk}, {Brett}, {Haldane}, {del R{\'\i}o},
  {Wiebe}, {Peterson}, {G{\'e}rard-Marchant}, {Sheppard}, {Reddy}, {Weckesser},
  {Abbasi}, {Gohlke}, \& {Oliphant}}]{Harris2020}
{Harris}, C.~R., {Millman}, K.~J., {van der Walt}, S.~J., {et~al.} 2020, \nat,
  585, 357, \dodoi{10.1038/s41586-020-2649-2}

\bibitem[{Heckman {et~al.}(1990)Heckman, Armus, \& Miley}]{Heckman1990}
Heckman, T.~M., Armus, L., \& Miley, G.~K. 1990, \apjs, 74, 833,
  \dodoi{10.1086/191522}

\bibitem[{{Hitomi Collaboration} {et~al.}(2016){Hitomi Collaboration},
  {Aharonian}, {Akamatsu}, {Akimoto}, {Allen}, {Anabuki}, {Angelini}, {Arnaud},
  {Audard}, {Awaki}, {Axelsson}, {Bamba}, {Bautz}, {Blandford}, {Brenneman},
  {Brown}, {Bulbul}, {Cackett}, {Chernyakova}, {Chiao}, {Coppi}, {Costantini},
  {de Plaa}, {den Herder}, {Done}, {Dotani}, {Ebisawa}, {Eckart}, {Enoto},
  {Ezoe}, {Fabian}, {Ferrigno}, {Foster}, {Fujimoto}, {Fukazawa}, {Furuzawa},
  {Galeazzi}, {Gallo}, {Gandhi}, {Giustini}, {Goldwurm}, {Gu}, {Guainazzi},
  {Haba}, {Hagino}, {Hamaguchi}, {Harrus}, {Hatsukade}, {Hayashi}, {Hayashi},
  {Hayashida}, {Hiraga}, {Hornschemeier}, {Hoshino}, {Hughes}, {Iizuka},
  {Inoue}, {Inoue}, {Ishibashi}, {Ishida}, {Ishikawa}, {Ishisaki}, {Itoh},
  {Iyomoto}, {Kaastra}, {Kallman}, {Kamae}, {Kara}, {Kataoka}, {Katsuda},
  {Katsuta}, {Kawaharada}, {Kawai}, {Kelley}, {Khangulyan}, {Kilbourne},
  {King}, {Kitaguchi}, {Kitamoto}, {Kitayama}, {Kohmura}, {Kokubun}, {Koyama},
  {Koyama}, {Kretschmar}, {Krimm}, {Kubota}, {Kunieda}, {Laurent}, {Lebrun},
  {Lee}, {Leutenegger}, {Limousin}, {Loewenstein}, {Long}, {Lumb}, {Madejski},
  {Maeda}, {Maier}, {Makishima}, {Markevitch}, {Matsumoto}, {Matsushita},
  {McCammon}, {McNamara}, {Mehdipour}, {Miller}, {Miller}, {Mineshige},
  {Mitsuda}, {Mitsuishi}, {Miyazawa}, {Mizuno}, {Mori}, {Mori}, {Moseley},
  {Mukai}, {Murakami}, {Murakami}, {Mushotzky}, {Nagino}, {Nakagawa},
  {Nakajima}, {Nakamori}, {Nakano}, {Nakashima}, {Nakazawa}, {Nobukawa},
  {Noda}, {Nomachi}, {O'Dell}, {Odaka}, {Ohashi}, {Ohno}, {Okajima}, {Ota},
  {Ozaki}, {Paerels}, {Paltani}, {Parmar}, {Petre}, {Pinto}, {Pohl}, {Porter},
  {Pottschmidt}, {Ramsey}, {Reynolds}, {Russell}, {Safi-Harb}, {Saito},
  {Sakai}, {Sameshima}, {Sato}, {Sato}, {Sato}, {Sawada}, {Schartel},
  {Serlemitsos}, {Seta}, {Shidatsu}, {Simionescu}, {Smith}, {Soong}, {Stawarz},
  {Sugawara}, {Sugita}, {Szymkowiak}, {Tajima}, {Takahashi}, {Takahashi},
  {Takeda}, {Takei}, {Tamagawa}, {Tamura}, {Tamura}, {Tanaka}, {Tanaka},
  {Tanaka}, {Tashiro}, {Tawara}, {Terada}, {Terashima}, {Tombesi}, {Tomida},
  {Tsuboi}, {Tsujimoto}, {Tsunemi}, {Tsuru}, {Uchida}, {Uchiyama}, {Uchiyama},
  {Ueda}, {Ueda}, {Ueno}, {Uno}, {Urry}, {Ursino}, {de Vries}, {Watanabe},
  {Werner}, {Wik}, {Wilkins}, {Williams}, {Yamada}, {Yamaguchi}, {Yamaoka},
  {Yamasaki}, {Yamauchi}, {Yamauchi}, {Yaqoob}, {Yatsu}, {Yonetoku}, {Yoshida},
  {Yuasa}, {Zhuravleva}, \& {Zoghbi}}]{Hitomi2016}
{Hitomi Collaboration}, {Aharonian}, F., {Akamatsu}, H., {et~al.} 2016, \nat,
  535, 117, \dodoi{10.1038/nature18627}

\bibitem[{{Ho} {et~al.}(2017){Ho}, {Martin}, {Kacprzak}, \&
  {Churchill}}]{Ho2017}
{Ho}, S.~H., {Martin}, C.~L., {Kacprzak}, G.~G., \& {Churchill}, C.~W. 2017,
  \apj, 835, 267, \dodoi{10.3847/1538-4357/835/2/267}

\bibitem[{{Huang} {et~al.}(2016){Huang}, {Chen}, {Johnson}, \&
  {Weiner}}]{Huang2016}
{Huang}, Y.-H., {Chen}, H.-W., {Johnson}, S.~D., \& {Weiner}, B.~J. 2016,
  \mnras, 455, 1713, \dodoi{10.1093/mnras/stv2327}

\bibitem[{{Hummels} {et~al.}(2019){Hummels}, {Smith}, {Hopkins}, {O'Shea},
  {Silvia}, {Werk}, {Lehner}, {Wise}, {Collins}, \& {Butsky}}]{Hummels2019}
{Hummels}, C.~B., {Smith}, B.~D., {Hopkins}, P.~F., {et~al.} 2019, \apj, 882,
  156, \dodoi{10.3847/1538-4357/ab378f}

\bibitem[{{Hunter}(2007)}]{Hunter2007}
{Hunter}, J.~D. 2007, Computing in Science and Engineering, 9, 90,
  \dodoi{10.1109/MCSE.2007.55}

\bibitem[{{Kere{\v s}} \& {Hernquist}(2009)}]{Keres2009}
{Kere{\v s}}, D., \& {Hernquist}, L. 2009, \apjl, 700, L1.
\newblock \doarXiv{0905.2186}

\bibitem[{{Kere{\v s}} {et~al.}(2005){Kere{\v s}}, {Katz}, {Weinberg}, \&
  {Dav{\'e}}}]{Keres2005}
{Kere{\v s}}, D., {Katz}, N., {Weinberg}, D.~H., \& {Dav{\'e}}, R. 2005,
  \mnras, 363, 2

\bibitem[{{Kritsuk} {et~al.}(2007){Kritsuk}, {Norman}, {Padoan}, \&
  {Wagner}}]{Kritsuk2007}
{Kritsuk}, A.~G., {Norman}, M.~L., {Padoan}, P., \& {Wagner}, R. 2007, \apj,
  665, 416, \dodoi{10.1086/519443}

\bibitem[{{Lochhaas} {et~al.}(2020){Lochhaas}, {Bryan}, {Li}, {Li}, \&
  {Fielding}}]{Lochhaas2020}
{Lochhaas}, C., {Bryan}, G.~L., {Li}, Y., {Li}, M., \& {Fielding}, D. 2020,
  \mnras, 493, 1461, \dodoi{10.1093/mnras/staa358}

\bibitem[{{Lochhaas} {et~al.}(2022){Lochhaas}, {Tumlinson}, {Peeples},
  {O'Shea}, {Werk}, {Simons}, {Juno}, {Kopenhafer}, {Augustin}, {Wright},
  {Acharyya}, \& {Smith}}]{Lochhaas2022}
{Lochhaas}, C., {Tumlinson}, J., {Peeples}, M.~S., {et~al.} 2022, arXiv
  e-prints, arXiv:2206.09925, \dodoi{10.48550/arXiv.2206.09925}

\bibitem[{Martin(2005)}]{Martin2005}
Martin, C.~L. 2005, \apj, 621, 227, \dodoi{10.1086/427277}

\bibitem[{Martin {et~al.}(2012)Martin, Shapley, Coil, Kornei, Bundy, Weiner,
  Noeske, \& Schiminovich}]{Martin2012}
Martin, C.~L., Shapley, A.~E., Coil, A.~L., {et~al.} 2012, \apj, 760, 127

\bibitem[{{McNamara} \& {Nulsen}(2007)}]{McNamara2007}
{McNamara}, B.~R., \& {Nulsen}, P.~E.~J. 2007, \araa, 45, 117,
  \dodoi{10.1146/annurev.astro.45.051806.110625}

\bibitem[{{Meiring} {et~al.}(2013){Meiring}, {Tripp}, {Werk}, {Howk},
  {Jenkins}, {Prochaska}, {Lehner}, \& {Sembach}}]{Meiring2013}
{Meiring}, J.~D., {Tripp}, T.~M., {Werk}, J.~K., {et~al.} 2013, \apj, 767, 49,
  \dodoi{10.1088/0004-637X/767/1/49}

\bibitem[{{Moe} {et~al.}(2009){Moe}, {Arav}, {Bautista}, \&
  {Korista}}]{Moe2009}
{Moe}, M., {Arav}, N., {Bautista}, M.~A., \& {Korista}, K.~T. 2009, \apj, 706,
  525, \dodoi{10.1088/0004-637X/706/1/525}

\bibitem[{{Morton}(2003)}]{Morton2003}
{Morton}, D.~C. 2003, \apjs, 149, 205, \dodoi{10.1086/377639}

\bibitem[{{Navarro} {et~al.}(1996){Navarro}, {Frenk}, \& {White}}]{Navarro1996}
{Navarro}, J.~F., {Frenk}, C.~S., \& {White}, S. D.~M. 1996, \apj, 462, 563,
  \dodoi{10.1086/177173}

\bibitem[{{Nicastro} {et~al.}(2023){Nicastro}, {Krongold}, {Fang},
  {Fraternali}, {Mathur}, {Bianchi}, {De Rosa}, {Piconcelli}, {Zappacosta},
  {Bischetti}, {Feruglio}, \& {Gupta}}]{Nicastro2023}
{Nicastro}, F., {Krongold}, Y., {Fang}, T., {et~al.} 2023, arXiv e-prints,
  arXiv:2302.04247, \dodoi{10.48550/arXiv.2302.04247}

\bibitem[{{Ogorzalek} {et~al.}(2017){Ogorzalek}, {Zhuravleva}, {Allen},
  {Pinto}, {Werner}, {Mantz}, {Canning}, {Fabian}, {Kaastra}, \& {de
  Plaa}}]{Ogorzalek2017}
{Ogorzalek}, A., {Zhuravleva}, I., {Allen}, S.~W., {et~al.} 2017, \mnras, 472,
  1659, \dodoi{10.1093/mnras/stx2030}

\bibitem[{{Osterman} {et~al.}(2011){Osterman}, {Green}, {Froning},
  {B{\'e}land}, {Burgh}, {France}, {Penton}, {Delker}, {Ebbets}, {Sahnow},
  {Bacinski}, {Kimble}, {Andrews}, {Wilkinson}, {McPhate}, {Siegmund}, {Ake},
  {Aloisi}, {Biagetti}, {Diaz}, {Dixon}, {Friedman}, {Ghavamian}, {Goudfrooij},
  {Hartig}, {Keyes}, {Lennon}, {Massa}, {Niemi}, {Oliveira}, {Osten},
  {Proffitt}, {Smith}, \& {Soderblom}}]{Osterman2011}
{Osterman}, S., {Green}, J., {Froning}, C., {et~al.} 2011, \apss, 335, 257,
  \dodoi{10.1007/s10509-011-0699-5}

\bibitem[{{Pan} \& {Scannapieco}(2010)}]{Pan2010}
{Pan}, L., \& {Scannapieco}, E. 2010, \apj, 721, 1765,
  \dodoi{10.1088/0004-637X/721/2/1765}

\bibitem[{{Pan} \& {Scannapieco}(2011)}]{Pan2011}
---. 2011, \pre, 83, 045302, \dodoi{10.1103/PhysRevE.83.045302}

\bibitem[{{Peeples} {et~al.}(2019){Peeples}, {Corlies}, {Tumlinson}, {O'Shea},
  {Lehner}, {O'Meara}, {Howk}, {Earl}, {Smith}, {Wise}, \&
  {Hummels}}]{Peeples2019}
{Peeples}, M.~S., {Corlies}, L., {Tumlinson}, J., {et~al.} 2019, \apj, 873,
  129, \dodoi{10.3847/1538-4357/ab0654}

\bibitem[{{Qu} {et~al.}(2022){Qu}, {Chen}, {Rudie}, {Zahedy}, {Johnson},
  {Boettcher}, {Cantalupo}, {Chen}, {Cooksey}, {DePalma},
  {Faucher-Gigu{\`e}re}, {Rauch}, {Schaye}, \& {Simcoe}}]{Qu2022}
{Qu}, Z., {Chen}, H.-W., {Rudie}, G.~C., {et~al.} 2022, \mnras, 516, 4882,
  \dodoi{10.1093/mnras/stac2528}

\bibitem[{{Rauch} {et~al.}(1996){Rauch}, {Sargent}, {Womble}, \&
  {Barlow}}]{Rauch1996}
{Rauch}, M., {Sargent}, W.~L.~W., {Womble}, D.~S., \& {Barlow}, T.~A. 1996,
  \apjl, 467, L5, \dodoi{10.1086/310187}

\bibitem[{{Reback} {et~al.}(2021){Reback}, {Jbrockmendel}, {McKinney}, {Van Den
  Bossche}, {Augspurger}, {Cloud}, {Hawkins}, {Roeschke}, {Gfyoung}, {Sinhrks},
  {Klein}, {Petersen}, {Hoefler}, {Tratner}, {She}, {Ayd}, {Naveh}, {Garcia},
  {Darbyshire}, {Schendel}, {Hayden}, {Shadrach}, {Saxton}, {Gorelli}, {Li},
  {Zeitlin}, {Jancauskas}, {McMaster}, {Battiston}, \& {Seabold}}]{Reback2021}
{Reback}, J., {Jbrockmendel}, {McKinney}, W., {et~al.} 2021,
  {pandas-dev/pandas: Pandas 1.3.5}, v1.3.5, Zenodo,  Zenodo,
  \dodoi{10.5281/zenodo.5774815}

\bibitem[{{Rubin} {et~al.}(2012){Rubin}, {Prochaska}, {Koo}, \&
  {Phillips}}]{Rubin2012}
{Rubin}, K.~H.~R., {Prochaska}, J.~X., {Koo}, D.~C., \& {Phillips}, A.~C. 2012,
  \apjl, 747, L26.
\newblock \doarXiv{1110.0837}

\bibitem[{{Rudie} {et~al.}(2019){Rudie}, {Steidel}, {Pettini}, {Trainor},
  {Strom}, {Hummels}, {Reddy}, \& {Shapley}}]{Rudie2019}
{Rudie}, G.~C., {Steidel}, C.~C., {Pettini}, M., {et~al.} 2019, \apj, 885, 61,
  \dodoi{10.3847/1538-4357/ab4255}

\bibitem[{{Rudie} {et~al.}(2012){Rudie}, {Steidel}, {Trainor}, {Rakic},
  {Bogosavljevi{\'c}}, {Pettini}, {Reddy}, {Shapley}, {Erb}, \&
  {Law}}]{Rudie2012}
{Rudie}, G.~C., {Steidel}, C.~C., {Trainor}, R.~F., {et~al.} 2012, \apj, 750,
  67, \dodoi{10.1088/0004-637X/750/1/67}

\bibitem[{Scannapieco(2017)}]{Scannapieco2017}
Scannapieco, E. 2017, \apj, 837, 28, \dodoi{10.3847/1538-4357/aa5d0d}

\bibitem[{Scannapieco \& Br{\"u}ggen(2015)}]{Scannapieco2015}
Scannapieco, E., \& Br{\"u}ggen, M. 2015, \apj, 805, 158,
  \dodoi{10.1088/0004-637X/805/2/158}

\bibitem[{Scannapieco \& Oh(2004)}]{Scannapieco2004}
Scannapieco, E., \& Oh, S.~P. 2004, The Astrophysical Journal, 608, 62–79,
  \dodoi{10.1086/386542}

\bibitem[{{Schmidt} {et~al.}(2009){Schmidt}, {Federrath}, {Hupp}, {Kern}, \&
  {Niemeyer}}]{Schmidt2009}
{Schmidt}, W., {Federrath}, C., {Hupp}, M., {Kern}, S., \& {Niemeyer}, J.~C.
  2009, \aap, 494, 127, \dodoi{10.1051/0004-6361:200809967}

\bibitem[{Steidel {et~al.}(2010)Steidel, Erb, Shapley, Pettini, Reddy,
  Bogosavljevi{\'c}, Rudie, \& Rakic}]{Steidel2010}
Steidel, C.~C., Erb, D.~K., Shapley, A.~E., {et~al.} 2010, \apj, 717, 289,
  \dodoi{10.1088/0004-637X/717/1/289}

\bibitem[{{Stewart} {et~al.}(2011){Stewart}, {Kaufmann}, {Bullock}, {Barton},
  {Maller}, {Diemand}, \& {Wadsley}}]{Stewart2011}
{Stewart}, K.~R., {Kaufmann}, T., {Bullock}, J.~S., {et~al.} 2011, \apj, 738,
  39.
\newblock \doarXiv{1103.4388}

\bibitem[{{Thacker} {et~al.}(2006){Thacker}, {Scannapieco}, \&
  {Couchman}}]{Thacker2006}
{Thacker}, R.~J., {Scannapieco}, E., \& {Couchman}, H.~M.~P. 2006, \apj, 653,
  86, \dodoi{10.1086/50865010.48550/arXiv.astro-ph/0606214}

\bibitem[{Thompson {et~al.}(2016)Thompson, Quataert, Zhang, \&
  Weinberg}]{Thompson2016}
Thompson, T.~A., Quataert, E., Zhang, D., \& Weinberg, D.~H. 2016, \mnras, 455,
  1830, \dodoi{10.1093/mnras/stv2428}

\bibitem[{{Tripp} {et~al.}(2011){Tripp}, {Meiring}, {Prochaska}, {Willmer},
  {Howk}, {Werk}, {Jenkins}, {Bowen}, {Lehner}, {Sembach}, {Thom}, \&
  {Tumlinson}}]{Tripp2011}
{Tripp}, T.~M., {Meiring}, J.~D., {Prochaska}, J.~X., {et~al.} 2011, Science,
  334, 952, \dodoi{10.1126/science.1209850}

\bibitem[{{Tumlinson} {et~al.}(2017){Tumlinson}, {Peeples}, \&
  {Werk}}]{Tumlinson2017}
{Tumlinson}, J., {Peeples}, M.~S., \& {Werk}, J.~K. 2017, \araa, 55, 389,
  \dodoi{10.1146/annurev-astro-091916-055240}

\bibitem[{{Tumlinson} {et~al.}(2013){Tumlinson}, {Thom}, {Werk}, {Prochaska},
  {Tripp}, {Katz}, {Dav{\'e}}, {Oppenheimer}, {Meiring}, {Ford}, {O'Meara},
  {Peeples}, {Sembach}, \& {Weinberg}}]{Tumlinson2013}
{Tumlinson}, J., {Thom}, C., {Werk}, J.~K., {et~al.} 2013, \apj, 777, 59,
  \dodoi{10.1088/0004-637X/777/1/59}

\bibitem[{{Turk} {et~al.}(2011){Turk}, {Smith}, {Oishi}, {Skory}, {Skillman},
  {Abel}, \& {Norman}}]{Turk2011}
{Turk}, M.~J., {Smith}, B.~D., {Oishi}, J.~S., {et~al.} 2011, \apjs, 192, 9,
  \dodoi{10.1088/0067-0049/192/1/9}

\bibitem[{{Uhlenbeck} \& {Ornstein}(1930)}]{Uhlenbeck1930}
{Uhlenbeck}, G.~E., \& {Ornstein}, L.~S. 1930, Physical Review, 36, 823,
  \dodoi{10.1103/PhysRev.36.823}

\bibitem[{{van de Voort} {et~al.}(2019){van de Voort}, {Springel}, {Mandelker},
  {van den Bosch}, \& {Pakmor}}]{vandeVoort2019}
{van de Voort}, F., {Springel}, V., {Mandelker}, N., {van den Bosch}, F.~C., \&
  {Pakmor}, R. 2019, \mnras, 482, L85, \dodoi{10.1093/mnrasl/sly190}

\bibitem[{Veilleux {et~al.}(2005)Veilleux, Cecil, \&
  Bland-Hawthorn}]{Veilleux2005}
Veilleux, S., Cecil, G., \& Bland-Hawthorn, J. 2005, \araa, 43, 769,
  \dodoi{10.1146/annurev.astro.43.072103.150610}

\bibitem[{{Voit} {et~al.}(2015){Voit}, {Bryan}, {O'Shea}, \&
  {Donahue}}]{Voit2015}
{Voit}, G.~M., {Bryan}, G.~L., {O'Shea}, B.~W., \& {Donahue}, M. 2015, \apjl,
  808, L30, \dodoi{10.1088/2041-8205/808/1/L30}

\bibitem[{{Werk} {et~al.}(2013){Werk}, {Prochaska}, {Thom}, {Tumlinson},
  {Tripp}, {O'Meara}, \& {Peeples}}]{Werk2013}
{Werk}, J.~K., {Prochaska}, J.~X., {Thom}, C., {et~al.} 2013, \apjs, 204, 17,
  \dodoi{10.1088/0067-0049/204/2/17}

\bibitem[{{Werk} {et~al.}(2014){Werk}, {Prochaska}, {Tumlinson}, {Peeples},
  {Tripp}, {Fox}, {Lehner}, {Thom}, {O'Meara}, {Ford}, {Bordoloi}, {Katz},
  {Tejos}, {Oppenheimer}, {Dav{\'e}}, \& {Weinberg}}]{Werk2014}
{Werk}, J.~K., {Prochaska}, J.~X., {Tumlinson}, J., {et~al.} 2014, \apj, 792,
  8, \dodoi{10.1088/0004-637X/792/1/8}

\bibitem[{{Werk} {et~al.}(2016){Werk}, {Prochaska}, {Cantalupo}, {Fox},
  {Oppenheimer}, {Tumlinson}, {Tripp}, {Lehner}, \& {McQuinn}}]{Werk2016}
{Werk}, J.~K., {Prochaska}, J.~X., {Cantalupo}, S., {et~al.} 2016, \apj, 833,
  54, \dodoi{10.3847/1538-4357/833/1/54}

\bibitem[{{Werner} {et~al.}(2009){Werner}, {Zhuravleva}, {Churazov},
  {Simionescu}, {Allen}, {Forman}, {Jones}, \& {Kaastra}}]{Werner2009}
{Werner}, N., {Zhuravleva}, I., {Churazov}, E., {et~al.} 2009, \mnras, 398, 23,
  \dodoi{10.1111/j.1365-2966.2009.14860.x}

\bibitem[{{Zahedy} {et~al.}(2021){Zahedy}, {Chen}, {Cooper}, {Boettcher},
  {Johnson}, {Rudie}, {Chen}, {Cantalupo}, {Cooksey}, {Faucher-Gigu{\`e}re},
  {Greene}, {Lopez}, {Mulchaey}, {Penton}, {Petitjean}, {Putman}, {Rafelski},
  {Rauch}, {Schaye}, {Simcoe}, \& {Walth}}]{Zahedy2021}
{Zahedy}, F.~S., {Chen}, H.-W., {Cooper}, T.~M., {et~al.} 2021, \mnras, 506,
  877, \dodoi{10.1093/mnras/stab1661}

\bibitem[{{Zhu} {et~al.}(2014){Zhu}, {M{\'e}nard}, {Bizyaev}, {Brewington},
  {Ebelke}, {Ho}, {Kinemuchi}, {Malanushenko}, {Malanushenko}, {Marchante},
  {More}, {Oravetz}, {Pan}, {Petitjean}, \& {Simmons}}]{Zhu2014}
{Zhu}, G., {M{\'e}nard}, B., {Bizyaev}, D., {et~al.} 2014, \mnras, 439, 3139,
  \dodoi{10.1093/mnras/stu186}

\bibitem[{{Zhuravleva} {et~al.}(2018){Zhuravleva}, {Allen}, {Mantz}, \&
  {Werner}}]{Zhuravleva2018}
{Zhuravleva}, I., {Allen}, S.~W., {Mantz}, A., \& {Werner}, N. 2018, \apj, 865,
  53, \dodoi{10.3847/1538-4357/aadae3}

\end{thebibliography}

\end{document}